\documentclass[journal]{IEEEtran}

\usepackage{amsmath,amsfonts}
\usepackage{algorithmic}
\usepackage{algorithm}
\usepackage{array}
\usepackage[caption=false,font=normalsize,labelfont=sf,textfont=sf]{subfig}
\usepackage{textcomp}
\usepackage{stfloats}
\usepackage{tabularx}
\usepackage{url}
\usepackage{verbatim}
\usepackage{graphicx}
\usepackage{cite}
\usepackage{amsmath,amsfonts}
\usepackage{algorithmic}
\usepackage{algorithm}
\usepackage{array}
\usepackage[caption=false,font=normalsize,labelfont=sf,textfont=sf]{subfig}
\usepackage{textcomp}
\usepackage{stfloats}
\usepackage{multirow} 
\usepackage{url}
\usepackage[table]{xcolor}
\definecolor{lightgray}{gray}{0.9}
\usepackage{verbatim}
\usepackage{graphicx}
\usepackage{booktabs}
\usepackage{array}  
\usepackage{hyperref}

\usepackage{xcolor} % 
\newif\ifshowrevs
\showrevstrue  
\ifshowrevs
  \newcommand{\rev}[1]{\textcolor{black}{#1}}
\else
  \newcommand{\rev}[1]{#1}
\fi

\begin{document}

\title{Evaluating the Usability of Microgestures for Text Editing Tasks in Virtual Reality}

\author{Xiang Li, Wei He, and Per Ola Kristensson

% <-this % stops a space
% \thanks{This paper was produced by the IEEE Publication Technology Group. They are in Piscataway, NJ.}% <-this % stops a space
% \thanks{Manuscript received \today.}
\thanks{Xiang Li and Per Ola Kristensson are with the Department of Engineering, University of Cambridge, Cambridge, United Kingdom.
E-mail: \{xl529, pok21\}@cam.ac.uk}
\thanks{Wei He is with the Thrust of Urban Governance and Design, The Hong Kong University of Science and Technology (Guangzhou), Guangzhou, China. 
E-mail: whe694@connect.hkust-gz.edu.cn}
}

% The paper headers
\markboth{IEEE TRANSACTIONS ON VISUALIZATION AND COMPUTER GRAPHICS,~Vol.~XX, No.~X, XXX~XXXX}%
{Li \MakeLowercase{\textit{et al.}}: Evaluating the Usability of Microgestures for Text Editing Tasks in Virtual Reality}

% \IEEEpubid{0000--0000/00\$00.00~\copyright~2024 IEEE}
% Remember, if you use this you must call \IEEEpubidadjcol in the second
% column for its text to clear the IEEEpubid mark.

\maketitle

\begin{abstract}
As virtual reality (VR) continues to evolve, traditional input methods such as handheld controllers and gesture systems often face challenges with precision, social accessibility, and user fatigue. \rev{These limitations motivate the exploration of microgestures, which promise more subtle, ergonomic, and device-free interactions.} We introduce microGEXT, a lightweight microgesture-based system designed for text editing in VR without external sensors, which utilizes small, subtle hand movements to reduce physical strain compared to standard gestures. We evaluated microGEXT in three user studies. In Study 1 ($N=20$), microGEXT reduced overall edit time and fatigue compared to a \rev{ray-casting + pinch menu baseline, the default text editing approach in commercial VR systems}. Study 2 ($N=20$) found that microGEXT performed well in short text selection tasks but was slower for longer text ranges. In Study 3 ($N=10$), participants found microGEXT intuitive for open-ended information-gathering tasks. Across all studies, microGEXT demonstrated enhanced user experience and reduced physical effort, offering a promising alternative to traditional VR text editing techniques.

\end{abstract}


%``ABC''
\section{Introduction}
As virtual reality (VR) continues to expand into various fields, the demand for effective input methods has become more critical than ever. Imagine a future where people rely on portable VR work environments, in such settings, prolonged use of controllers or gestures—whether through handheld devices or body movements—poses significant challenges~\cite{grubert2018office,biener2023extended}, particularly regarding fatigue, such as the well-known ``gorilla arm effect''~\cite{palmeira2024quantifying,hincapi2014consumed}. Furthermore, in confined spaces, such as airplanes~\cite{williamson2019planevr} or buses~\cite{tseng2023fingermapper}, using large-scale whole-body movements~\cite{floyd2021limited,xu_dmove_2019,li2024onbodymenu,mueller2023towards} or extending the arms~\cite{xu_evaluation_2022,li2021vrcaptcha} for interaction could either disturb others or be entirely impractical due to spatial constraints~\cite{yang2018sharespace}. These limitations can seriously hinder the effectiveness of VR input methods in real-world scenarios.

The \emph{gorilla arm} effect, common during VR text input, arises from extended arm positions used for virtual keyboard interaction, increasing torque on the shoulder and elbow joints—3.77 times more at the shoulder and twice as much at the elbow compared to relaxed arm positions~\cite{palmeira2024quantifying}. Such postures cause fatigue, making tasks like character selection or text editing cumbersome. While large-scale gestures exacerbate this issue, \rev{microgestures offer a promising alternative: they are subtle, socially unobtrusive, and physically less demanding, while retaining the benefits of gesture-based interaction~\cite{chan_user_2016,xu2020results}.} These subtle, minimal movements reduce physical strain and are well-suited for tasks requiring fine control. 

To address these challenges, we introduce \emph{microGEXT}, a microgesture-based system for text editing in VR. Unlike text entry, text editing benefits more from shortcuts~\cite{le_shortcut_2020} that enable efficient structured interactions. Our microGEXT system leverages built-in VR cameras to detect small, ergonomic movements without external hardware, enabling precise control for tasks such as caret navigation and text selection~\cite{hu_evaluation_2022} while minimizing fatigue. \rev{For comparison, we evaluate microGEXT against a ray-casting + pinch menu baseline, which is the commercial default method for VR text editing in current head-mounted displays.}

We evaluate the usability of microGEXT for VR text editing tasks in three studies. In Study 1 ($N=20$), participants performed common text editing tasks (e.g., navigation, selection, copy-paste) to compare microGEXT with the baseline. Results showed no significant difference in overall edit times, with microGEXT being faster in some tasks. Participants also reported improved efficiency and satisfaction. Building on this, Study 2 ($N=20$) focused on microGEXT’s accuracy and speed in precise text range selection, such as highlighting characters or paragraphs. While microGEXT matched the Baseline for shorter selections, it required more time for longer ones but was rated as smoother, less demanding, and less frustrating overall. It was also ranked significantly easier to use, higher in presence, and more preferred, while notably reducing perceived fatigue compared to the Baseline. Finally, Study 3 ($N=10$) explored microGEXT in open-ended information-gathering tasks, where participants selected, copied, and pasted data between a web browser and a note-taking app in VR. Feedback highlighted microGEXT’s intuitive, efficient, and fatigue-free performance, particularly in enabling seamless task switching during extended sessions.

In summary, our work makes the following contributions:

\begin{itemize}
\item We introduce \textbf{microGEXT}, a lightweight microgesture-based framework that facilitates precise and efficient text editing in VR, achieving high recognition accuracy using built-in cameras, without external sensors.
\item We demonstrate that microGEXT significantly reduces edit times for commands like \textsc{Cut}, \textsc{Delete}, and \textsc{Select All}, while lowering physical demand, mental effort, and frustration. It performs well across diverse text range selection tasks, though less effectively for extended ranges, and delivers higher user satisfaction and lower fatigue in structured and open-ended VR text editing scenarios.
\end{itemize}

\begin{figure*}
\centering
\includegraphics[width=\linewidth]{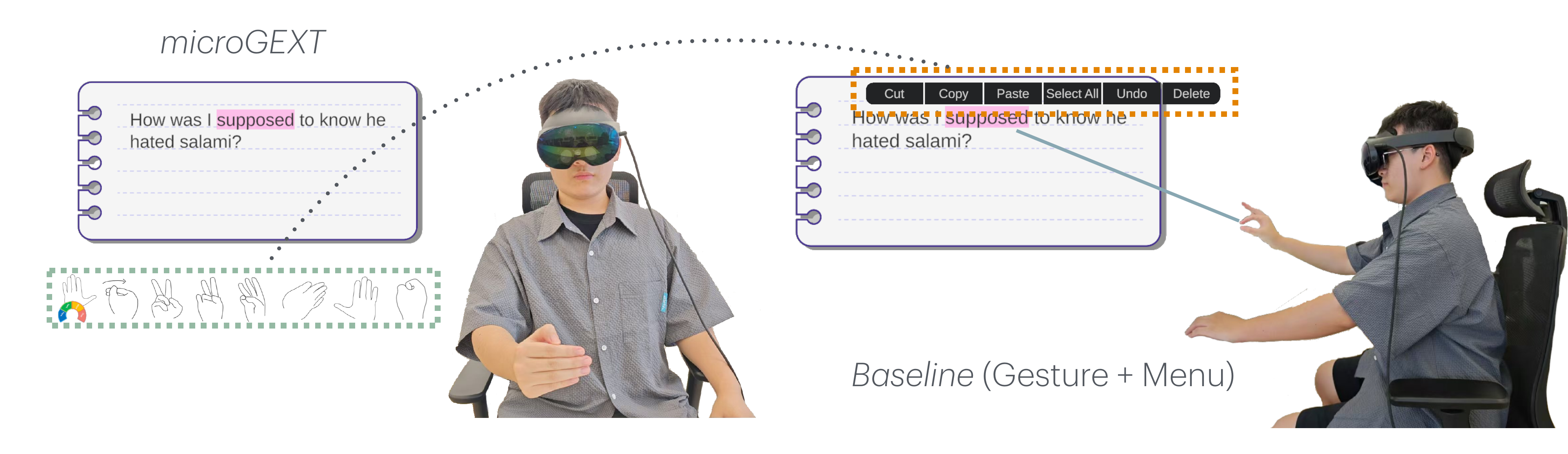}
\caption{In this paper, we introduce the microGEXT system, which replaces traditional menu-based interactions with intuitive microgestures for text editing tasks in virtual environments. This teaser figure highlights the contrast between the conventional text editing menu and the corresponding microgestures used for functions such as mode switching, caret navigation, range selection, cut, copy, paste, select all, undo, and delete (from left to right).}
\label{fig:teaser}
\end{figure*}

\section{Related Work}

\subsection{Text Editing in Real World and Virtual Environments}

Text selection and editing in immersive virtual environments (VEs) present unique challenges due to the spatial constraints and limitations of traditional input devices like keyboards and mice. Song et al. explored these challenges, highlighting the inefficiencies of conventional devices in VEs and proposing controller-based systems as an alternative to improve text selection performance~\cite{song_exploring_2024}. De Rosa et al. further addressed the spatial interaction difficulties inherent in VEs, focusing on precision-driven tools for text editing~\cite{de_rosa_arrow2edit_2023}.

\rev{A broad line of research has therefore explored alternative interaction techniques tailored to immersive contexts. Directional motion-based text selection systems (e.g., DMove~\cite{xu_dmove_2019}) reduce reliance on 2D metaphors by mapping gestures to navigation. Hybrid approaches integrate eye-tracking and gestures, such as M[eye]cro~\cite{wambecke_meyecro_2021}, to enhance speed and accuracy. Other systems augment text selection with novel sensing modalities, including spatial gestures~\cite{dudley_fast_2018}), radar-based tracking~\cite{hajika_radarhand_2024}), and tactile augmentation~\cite{kim_vibaware_2023}. Together, these works demonstrate that immersive text editing requires interaction techniques that are precise, ergonomic, and compatible with the physical and social constraints of VR. However, many solutions still rely on large movements or external hardware, which can induce fatigue and limit portability. This gap motivates exploring microgestures as a lightweight and ergonomic alternative.}

\subsection{(Micro)Gestures for Interaction}

Gestures have emerged as a promising alternative to conventional input devices, particularly in scenarios where traditional methods are impractical. Sellier et al. demonstrated the natural interaction advantages of gesture interfaces for digital manipulation~\cite{sellier_evaluating_2024}, while Wobbrock et al. highlighted the value of user-defined gestures for tailoring interactions to individual preferences, especially on touch surfaces~\cite{wobbrock_user-defined_2009}. \rev{For text editing specifically, Le et al. showed that shortcut gestures can significantly improve workflow efficiency~\cite{le_shortcut_2020}, and multimodal approaches (e.g., gaze–hand combinations~\cite{slambekova_gaze_2012}) further enhance precision. Beyond optical tracking, sonar-based systems such as FingerIO~\cite{nandakumar_fingerio_2016} illustrate the technical diversity of gesture sensing.}

\rev{Microgestures extend this line of work by explicitly targeting fatigue reduction and social acceptability in constrained spaces. Chan et al.~\cite{chan_user_2016} elicited user-preferred microgestures, highlighting their intuitiveness and minimal effort. Subsequent studies demonstrated effective applications: Faisandaz et al.~\cite{faisandaz_get_2023} examined eyes-free use, Sharma et al. explored grasping microgestures~\cite{sharma_grasping_2019} and one-handed text editing~\cite{sharma_solofinger_2021}), and Kandoi et al.~\cite{kandoi_intentional_2023} emphasized intentionality for high-precision tasks. Additional designs explored rhythmic patterns~\cite{freeman_rhythmic_2017}, one-hand use in constrained spaces~\cite{soliman_fingerinput_2018}, and mobile contexts such as biking~\cite{tan_bikegesture_2017}. Frameworks like iFAD further illustrate the potential of microgestures for multitasking~\cite{vatavu_ifad_2023}.}

\rev{Beyond application prototypes, recent elicitation and systematic studies provide methodological foundations for microgesture design. Chaffangeon Caillet et al.~\cite{chaffangeon_caillet_microgesture_2025} identified transferable microgestures across grasp contexts, Villarreal-Narvaez et al.~\cite{villarreal-narvaez_brave_2024} synthesized 267 elicitation studies into a taxonomy, and Gheran et al.~\cite{gheran_repliges_2022} introduced tools for replication and consolidation. These works stress that gesture sets must be justified not only ergonomically but also conceptually, with attention to context and task mappings. Our work builds directly on these insights by designing and evaluating microgestures specifically for VR text editing, a domain that has received limited attention despite its importance for productivity in immersive environments.}

\section{microGEXT: A Microgesture Recognition Framework for Text Editing}

Previous studies have highlighted the benefits of gestures and microgestures in immersive interactions, but a fully intuitive text editing experience without external cameras, wearables, or specialized hardware remains undeveloped. Many existing solutions are limited by hardware demands or induce fatigue from prolonged mid-air hand use. Our research aims to develop a lightweight VR text editing system that replaces large body movements with subtle microgestures, reducing physical strain. To facilitate further research, we release our open-source code and anonymized datasets.

\subsection{Gesture Dataset}

We used the Meta Quest Pro VR headset with the XR Hand package\footnote{\href{https://docs.unity3d.com/Packages/com.unity.xr.hands@1.4/manual/}{https://docs.unity3d.com/Packages/com.unity.xr.hands@1.4/manual/}} to capture hand skeleton data for gesture recognition. \rev{Our gesture design was informed by prior elicitation work. In particular, Chan et al.~\cite{chan_user_2016} proposed a dataset of user-defined single-hand microgestures, identifying prevalent conceptual themes such as directional swipes, finger touches, and static hand poses. From this set we selected gestures that (i) were rated as comfortable and memorable in prior studies, (ii) mapped naturally to common text editing commands (e.g., \textsc{Cut}, \textsc{Paste}, \textsc{Select All}), and (iii) could be reliably recognized using only the Quest Pro’s built-in cameras. Three gestures (\textsc{Pinch}, \textsc{Circle}, and dynamic \textsc{Swipe}) were newly designed to extend command coverage. For each mapping, we provide a rationale in Section~\ref{choice}, ensuring semantic alignment with the assigned editing command.}

Ten participants were recruited to interact with a text-editing application, contextualizing each gesture. Before data collection, participants viewed a demonstration video and then performed each gesture 20 times. For static gestures, data was clipped to a standard 2-second duration, while dynamic gestures were clipped to capture complete movement, averaging 5 seconds per gesture. Data was recorded at the Quest Pro’s native frame rate of 72 Hz. 

\rev{To improve recognition, we adopted a multi-state segmentation for the dynamic \textsc{Swipe} gesture. Each clip of length \textit{T} was divided into sub-states (0–3), automatically labeled based on finger position: smaller values denoted positions near the \textsc{IndexTip}, while larger values indicated the \textsc{IndexDistal}. Other gestures were assigned a single sub-state label ‘4’. This design enabled finer phase-level classification, improving robustness.} We also included a dedicated \textit{Null} class, capturing non-intentional hand movements (e.g., resting pose, incidental pinches). \rev{This is critical for delimiter-free recognition, reducing false activations when users are not performing deliberate gestures, as emphasized in prior research~\cite{faisandaz_get_2023,song_hotgestures_2023}}.

Participants synchronized their swipe with a visual clock in VR, sliding their index finger from 0\% to 100\% and back within 5 seconds, allowing for automatic sub-state labeling. Data collection lasted approximately 2 seconds per static gesture and 5 seconds for dynamic gestures.

\subsection{Model Architecture}

\begin{figure*}[t]
    \centering
    \includegraphics[width=\linewidth]{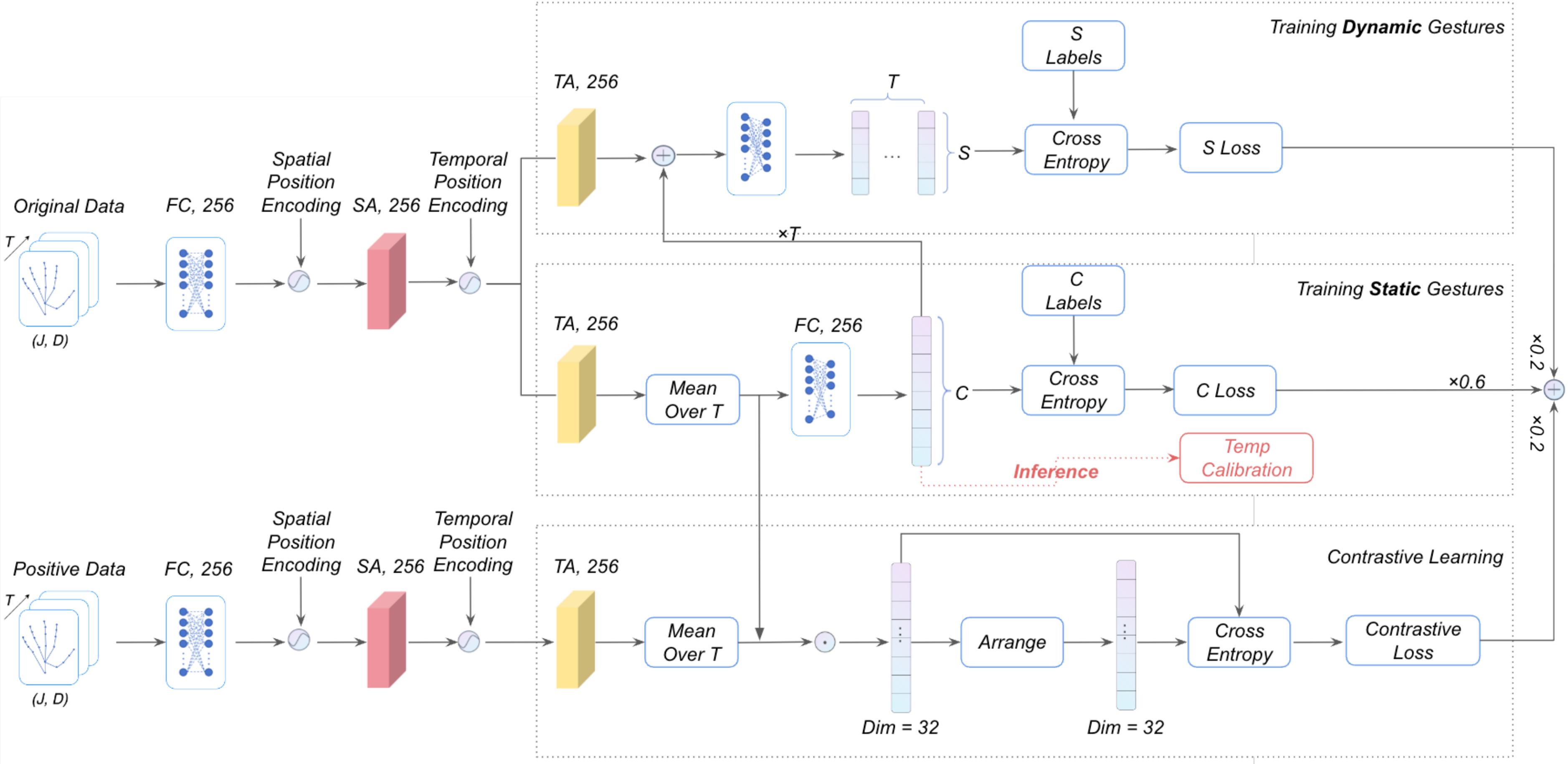}
    \caption{Overview of the recognition network and framework with contrastive learning. The network consists of Spatial and Temporal Position Encoding, Temporal Attention (TA), and Fully Connected (FC) layers (with ReLU activation function and layer-normalization). The model processes dynamic gestures using a sub-state prediction branch and static gestures using a classification branch. Contrastive learning is applied to enhance the model’s ability to distinguish between gestures. The total loss is a weighted combination of dynamic, static, and contrastive losses. Numbers in each block represent the output dimensions.The temperature scaling layer was added before the final prediction output in the calibration process.}
    \label{fig:model}
\end{figure*}

For our uni-manual gesture recognition study, we adapted the HotGestures multi-task deep learning architecture~\cite{song_hotgestures_2023}, known for its effectiveness in static and dynamic gesture recognition. As shown in Figure~\ref{fig:model}, the model processes hand skeleton sequences with advanced spatial and temporal encoding to handle various gesture types. It uses hand skeleton graphs with dimensions (\textit{J, D}), where \textit{J} is the number of joints, and \textit{D} is the feature dimensions per joint. The data passes through a fully connected layer (256 units) for feature extraction, followed by a Spatial Position Encoding (256 units) module for joint relationships and a Temporal Position Encoding module for time-based information. A Temporal Attention (256 units) block highlights key temporal segments, which are crucial for distinguishing dynamic gestures and filtering irrelevant frames in static ones. A temperature scaling layer is utilized before the output to enhance model calibration, improving confidence in predictions~\cite{guo2017calibration}.

\subsubsection{Dynamic Gesture Recognition} Our model uses a specialized pipeline for dynamic gesture recognition. After the TA layer, the extracted features are concatenated with those from the initial fully connected layer. This combined representation is passed through another fully connected layer (FC, 5), followed by \textit{T} parallel branches, where \textit{T} is the number of frames in the sequence. Each branch outputs a sub-state probability \textbf{S}, allowing frame-by-frame analysis of dynamic gestures. The cross-entropy loss for this branch is computed and scaled by a factor of 0.2.

\subsubsection{Static Gesture Recognition} For static gesture recognition, the output of the TA layer is pooled across the temporal dimension using mean pooling, followed by a fully connected layer (FC, 8) to produce the final classification output \textbf{C}. This branch also applies cross-entropy loss, scaled by 0.6.

\subsubsection{Contrastive Learning} We introduced incorporated contrastive learning to enhance the original model’s discriminative capabilities~\cite{tian2020makes}. Positive gesture samples follow a similar pipeline as the static gesture branch, with the addition of a feature arrangement step before applying a contrastive loss function. This contrastive learning component, weighted by 0.2, significantly helps the model learn more robust features by contrasting similar and dissimilar gesture samples. 

Finally, the overall loss for the model is a weighted combination of dynamic gesture loss, static gesture loss, and contrastive loss, ensuring a balanced optimization objective that accommodates the diverse nature of gesture recognition.

\begin{figure*}[ht]
    \centering
    \includegraphics[width=0.9\linewidth]{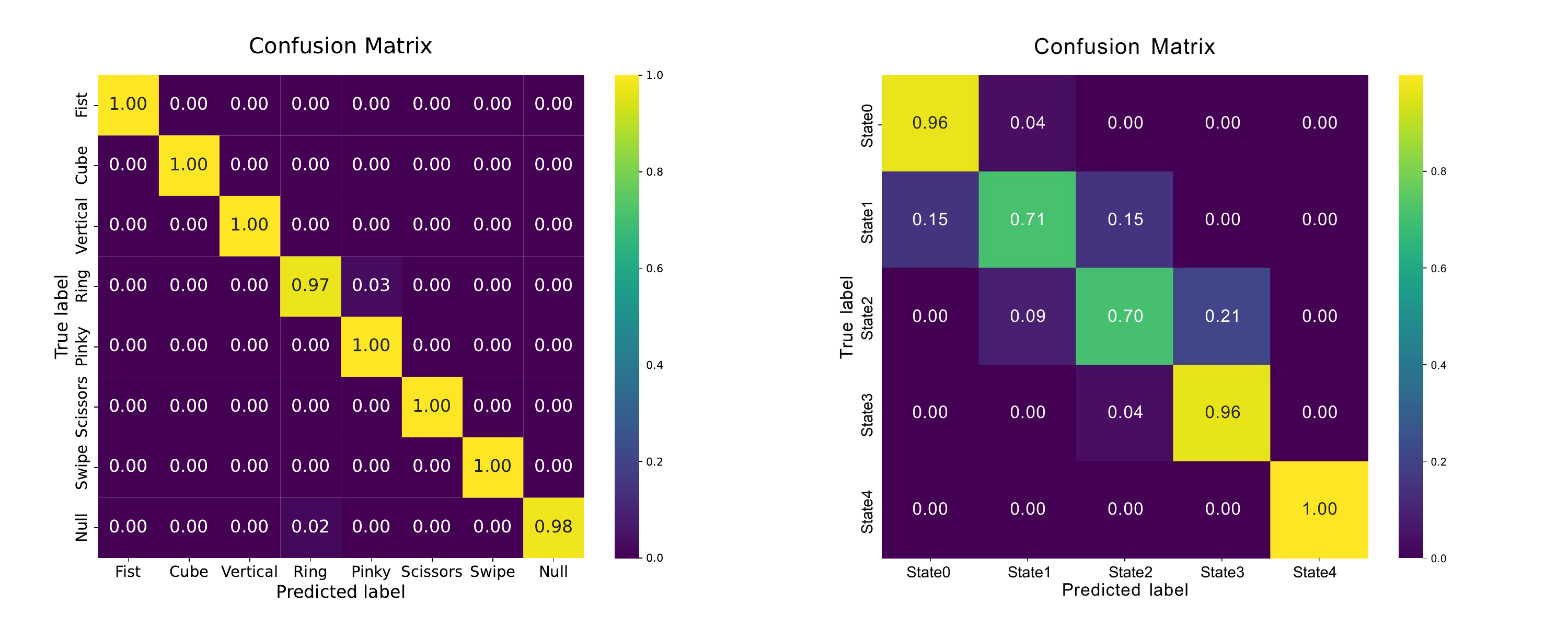}
    \caption{Class and State Confusion Matrices for gesture recognition performance. The Class Confusion Matrix (left) shows the classification accuracy for each gesture class, with high accuracy across most classes, though some minor misclassifications are observed for gestures like ``Null'' and ``Ring''. The State Confusion Matrix (right) displays the accuracy of sub-state predictions for the ``Swipe'' gesture, with good performance in most states, though state transitions show some misclassification between adjacent sub-states.}
    % \Description{Two confusion matrices: the Class Confusion Matrix (left) and State Confusion Matrix (right). The Class matrix shows high accuracy for gestures like ‘Fist,’ ‘Open,’ and ‘Swipe,’ while minor misclassifications occur in gestures like ‘Ring’ and ‘Scissors.’ The State matrix focuses on sub-state prediction within the ‘Swipe’ gesture. Both matrices use color bars to represent prediction accuracy from low (purple) to high (yellow).}
    \label{fig:confustion_matrix}
\end{figure*}

\subsection{Training and Implementation Details}

We set the window size to \textit{T} = 20 and used \textit{J} = 11 joints (all fingertips, one joint below each fingertip, and the wrist root) as suggested by Song et al.~\cite{song_hotgestures_2023}. The feature dimension was set to \textit{D} = 7, representing the relative 3D position and 4D quaternion rotation of 10 joints relative to the wrist. For training and evaluation, we selected one participant’s data as the test set and used the remaining data for training, employing cross-validation across participant combinations. The training was conducted using PyTorch on a system with an Nvidia GeForce RTX 4090 GPU and an Intel Core i9-13900K CPU.

We used the Adam optimizer with a learning rate of 0.1\%, a learning rate scheduler that reduces on a plateau with patience of 5 epochs, and a batch size of 32. Null gestures were included in the training to prevent false positives. The model loss function combined three Cross-Entropy (CE) losses: one for gesture classification, one for sub-state prediction, and one for contrastive learning, represented by $\mathcal{L}{\text{class}}$, $\mathcal{L}{\text{state}}$, and $\mathcal{L}_{\text{contrastive}}$. The loss weights were $\alpha = 0.6$, $\beta = 0.2$, and $\gamma = 0.2$, corresponding to classification, sub-state, and contrastive tasks, respectively.

\begin{equation}
\mathcal{L}_{\text{total}} = \alpha \times \mathcal{L}_{\text{class}} + \beta \times \mathcal{L}_{\text{state}} + \gamma \times \mathcal{L}_{\text{contrastive}}
\label{eq:total_loss}
\end{equation}

\subsubsection{Offline Gesture Recognition}

Figure \ref{fig:confustion_matrix} shows the confusion matrices for the window-level class prediction and sub-state prediction on the validation set. For the class prediction, all gesture recognition accuracy is above 97\%, indicating the good performance of our recognition model. ``Null'' and ``Ring'' gestures have relatively low accuracy since the ``Ring'' gesture is as easy to detect as the ``Pinky'' gesture, and the ``Null'' gesture is similar to the ``Cube'' gesture. Note that other gestures have 100\% accuracy which means the model overfit. For the state prediction, the state `4' has the highest accuracy, and states `0' and `3' are higher than states `1-2'. This is expected because class loss weight $\mathcal{L}_{\text{class}}$ is higher than state loss weight $\mathcal{L}_{\text{state}}$, forcing the model to learn better at class prediction and be less sensitive about sub-states. And the 0\% and 100\% are much easier to recognize than other sub-states.

\subsubsection{Online Gesture Recognition}

During the online gesture recognition phase, we implemented a dynamic windowing approach for processing the continuous data stream, rather than relying on a static window size. This allows for real-time gesture detection in virtual reality, where continuous and dynamic input is crucial. A data buffer is maintained, storing the most recent \( T \) frames from the stream. These frames are used as input to predict both the gesture class \textbf{C} and the corresponding state sequence \textbf{S}.

To enhance accuracy and minimize false recognitions during real-time detection, we incorporated a finite-state machine (FSM), as suggested by Song et al.~\cite{song_hotgestures_2023}. The FSM governs the gesture recognition process through three distinct states. Initially, the FSM is in state \( S1 \), where no gestures are detected. When a gesture is identified with a class probability  \textbf{C}  above a predefined threshold \( \delta \), the FSM transitions to state \( S2 \). In state \( S2 \), if the same gesture class is consistently detected over \( N \) consecutive frames with probabilities exceeding \( \delta \), the FSM moves to state \( S3 \), finalizes the gesture recognition, and returns to \( S1 \). Conversely, if a different gesture class is detected in \( S2 \) with a probability greater than \( \delta \), the FSM stays in \( S2 \) without transitioning to \( S3 \). If at any point the model’s predicted class probability drops below \( \delta \), the FSM immediately reverts to \( S1 \). This system ensures that only gestures with high confidence are processed, effectively reducing false positives and improving the robustness of real-time interaction. We implemented our FSM in Unity and fine-tuned the parameters \( N \) and \( \delta \) based on the results of an in-lab formative study with four participants. During the formative study, they were asked to perform each gesture sequentially to test the speed and accuracy of the model. After completing all the gesture commands, participants were required to give their feedback about the model detection. By experimenting with different parameter values, we found \(N=10\) and \(\delta = 0.95\) yield the best user experience and lowest error rate.

\subsection{Gestures and Features}
\label{choice}

Our microGEXT system utilizes a set of microgestures designed to facilitate intuitive and efficient text editing in virtual environments. These gestures are performed with the dominant hand, while the non-dominant hand controls range selection modes~\cite{li2023swarm,li2024swarm}. The following sections detail the interaction flow of our selected microgestures and their corresponding features.

\subsubsection{Cut (Scissor Gesture)} The \textit{Scissor} gesture mimics a cutting motion with the index and middle fingers, activating the \textsc{Cut} function to remove selected text and store it in the clipboard. Users select text, perform the gesture, and the text is cut.

\subsubsection{Copy (Ring Gesture)} The \textit{Ring} gesture, formed by connecting the index finger and thumb, triggers the \textsc{Copy} function to copy selected text without removing it. This efficient gesture enables copying without external controllers.

\subsubsection{Caret Navigation and Range Selection (Swipe Gesture)}
\rev{Newly designed for this work, inspired by DigitSpace~\cite{huang_digitspace_2016} and PinchWatch~\cite{loclair2010pinchwatch},} the \textit{Swipe} gesture slides the thumb along the index finger, offering:
\begin{itemize}
\item \textbf{Caret Navigation:} A simple swipe moves the caret within text for precise positioning.
\item \textbf{Range Selection:} With a long press, the swipe highlights text, allowing efficient selection.
\end{itemize}

\subsubsection{Undo (Open Gesture)} The \textit{Open} gesture, represented by an open palm, activates \textsc{Undo} to reverse the last action, enabling quick corrections.

\subsubsection{Delete (Fist Gesture)} The \textit{Fist} gesture closes the hand into a fist to activate \textsc{Delete}, removing selected text without copying it to the clipboard. This straightforward gesture facilitates quick text deletion.

\subsubsection{Select All (Vertical Gesture)}
The \textit{Vertical} gesture, performed by raising all five fingers, triggers \textsc{Select All}, allowing users to highlight all text in a document efficiently.

\subsubsection{Paste (Pinky Gesture)}
Extending the pinky while folding other fingers triggers the \textsc{Paste} function, enabling users to quickly insert it at the caret’s position.

\subsubsection{Range Selection Mode (Wrist Rotation)}
\rev{Inspired by prior VR 3D modeling work~\cite{song_exploring_2024} and efficient special
character entry~\cite{song_efficient_2022},} the left hand controls \textit{Range Selection Mode} through wrist rotation, cycling between character, word, sentence, and paragraph selection. This intuitive interaction adjusts selection granularity without disrupting workflow.

\begin{figure}[t]
    \centering
    \includegraphics[width=\linewidth]{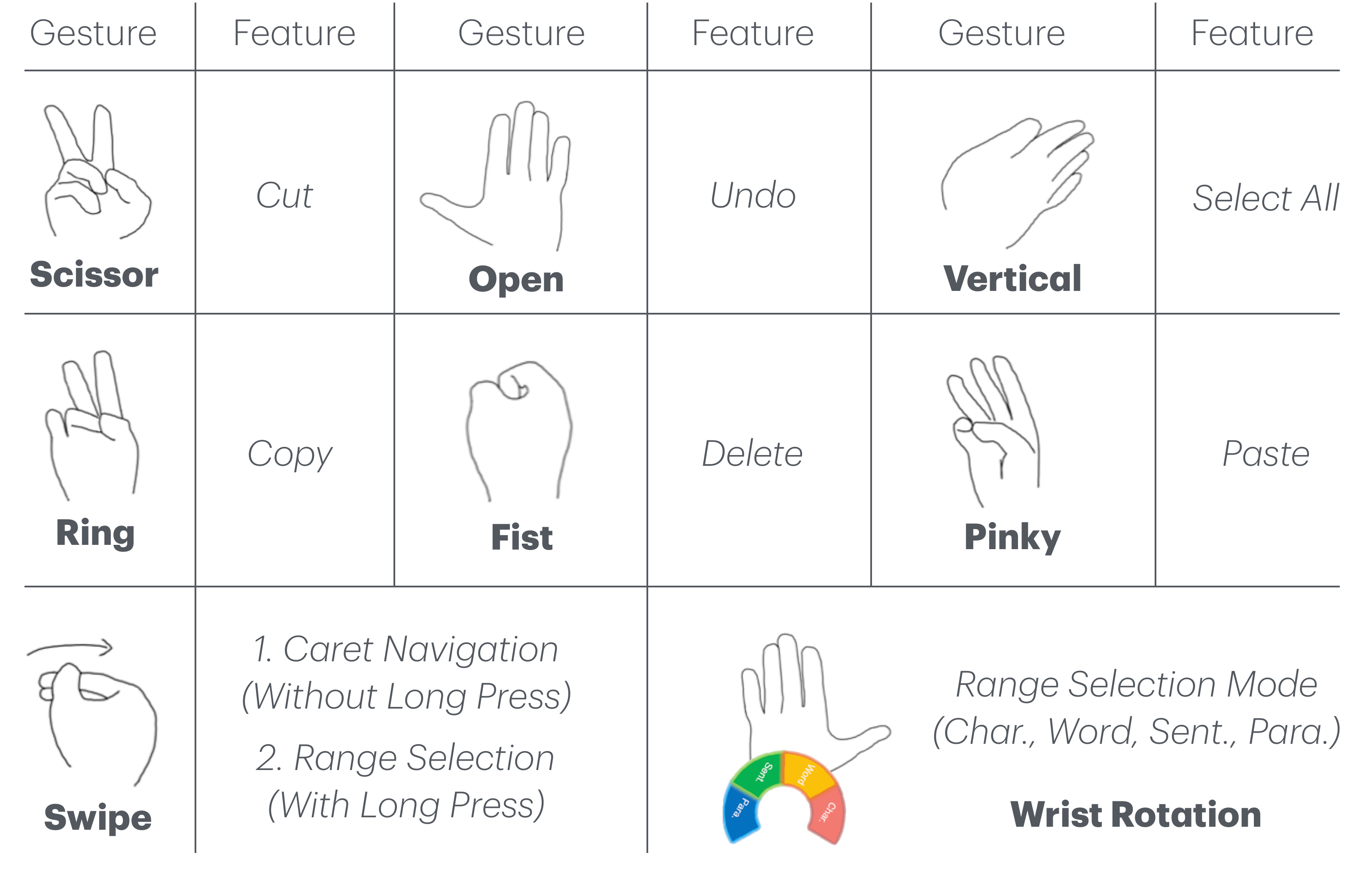}
    \caption{Our microGEXT’s microgestures for text editing tasks. The right hand (dominant hand) performs gestures such as Scissor (Cut), Ring (Copy), Swipe (Caret Navigation and Range Selection), Open (Undo), Fist (Delete), Vertical (Select All), and Pinky (Paste). The left hand (non-dominant hand) is used for wrist rotation to control the range selection mode, allowing the user to switch between selecting characters, words, sentences, and paragraphs.}
    % \Description{A table showing microGEXT's hand gestures for text editing tasks. Right-hand gestures include Scissor (Cut), Ring (Copy), Swipe (Caret Navigation and Range Selection), Open (Undo), Fist (Delete), Vertical (Select All), and Pinky (Paste). The left hand uses wrist rotation to switch between range modes such as character, word, sentence, and paragraph selection.}
    \label{fig:system_feature}
\end{figure}

\section{User Study 1: Comparing microGEXT and Baseline Conditions for Text Editing Command}

Study 1 investigates the feasibility of using microgestures for text editing tasks in virtual environments. We conducted a within-subjects user study ($N=20$), evaluating the impact of microGEXT on usability (e.g., user experience, system usability, and perceived workload) and utility (i.e., edit times and reattempts). The microGEXT was compared to a Baseline condition (see Section \ref{sec:Baseline}) to assess overall effectiveness.

While the lightweight framework for microgesture detection is robust, prior work indicates that gesture-based input can face challenges such as limited memorability and occasional recognition errors, which may reduce potential efficiency gains~\cite{kristensson2014inviscid}. \rev{We therefore examine whether microGEXT achieves comparable speed and accuracy to the Baseline condition (\textbf{H1}). At the same time, because microgestures involve minimal physical effort, they may be perceived as less fatiguing and more comfortable for extended use, which could translate into higher ratings of usability and user experience compared to the Baseline (\textbf{H2}).}

\subsection{Method}

\subsubsection{Baseline}
\label{sec:Baseline}

In comparison to the microGEXT system, we selected the text editing tool selection method from the OpenXR\footnote{\href{https://mbucchia.github.io/OpenXR-Toolkit/}{https://mbucchia.github.io/OpenXR-Toolkit/}} package as the Baseline condition, representing a conventional and well-established approach for tool selection in text editing tasks. This Baseline used a \rev{ray-casting + pinch menu system, reflecting the default design in many commercial VR applications (e.g., Oculus Browser, Immersed)}. Participants pointed with ray casting to select the desired tool (e.g., \textsc{Copy}, \textsc{Paste}, \textsc{Cut}) and then confirmed with a pinch gesture. \rev{We chose this Baseline because it is the most common default for text editing and command activation in commercial VR systems, thereby providing a familiar and practical comparator rather than a novel or arbitrary technique.} The menu interface design adhered closely to the default OpenXR sample menus, ensuring consistency.

\subsubsection{Participants and Apparatus} 
\label{sec:Participants}

A total of 20 participants (13 male, 7 female) were recruited from a local university. The age range of the participants was between 17 and 28 years ($M = 23.25, SD = 2.71$). All participants were students, right-handed, and reported previous VR experience (familiarity 1–7, $M = 4.25, SD = 1.71$). None of the participants involved in data collection participated in the user study. The virtual environment was provided via a Meta Quest Pro VR HMD with hand tracking enabled for interaction. The program was developed in Unity Engine (version 2022.3.3f1c1) with the OpenXR Hand package (version 1.4.1). The HMD was connected to a high-performance computer via Quest Link, equipped with an Intel i9-13900K CPU, an NVIDIA GeForce RTX 4090 GPU, and 64GB RAM. \rev{The study protocol was approved by the Cambridge University Engineering Department Ethics Committee (\#452), and all participants provided informed consent.}

\subsubsection{Procedure}
\label{sec:procedure}

Before the experiment, participants signed a consent form and completed a demographic questionnaire. They were then shown a video demonstration introducing each of the seven gestures. Following this, participants viewed video clips of the gestures associated with each tool, as described in the previous section, and were instructed to memorize them as much as possible. Participants were then allowed to practice all eight gestures with feedback provided by the gesture recognition system. No data were recorded during this practice phase, which ensured that participants fully understood how to use both the Baseline and microGEXT systems for text editing tasks. If a participant was unable to complete the practice session successfully, they would not have been allowed to proceed with the formal study or receive compensation. The formal experiment followed, with participants performing tasks under two conditions: Baseline and microGEXT. The order of these conditions was counterbalanced across participants using a Latin square design. The experiment consisted of instruction-based random tasks, where participants used text editing tools without selecting specific text snippets. After each condition, participants were given a break and asked to complete a questionnaire assessing their experience. Consequently, the overall study comprised a total of 1,600 trials, calculated as 2 (Condition) $\times$ 8 (Command) $\times$ 5 (Round) $\times$ 20 (Participant). On average, the entire study took approximately 40 minutes per participant, and participants received compensation.

\begin{figure}[t]
    \centering
    \includegraphics[width=\linewidth]{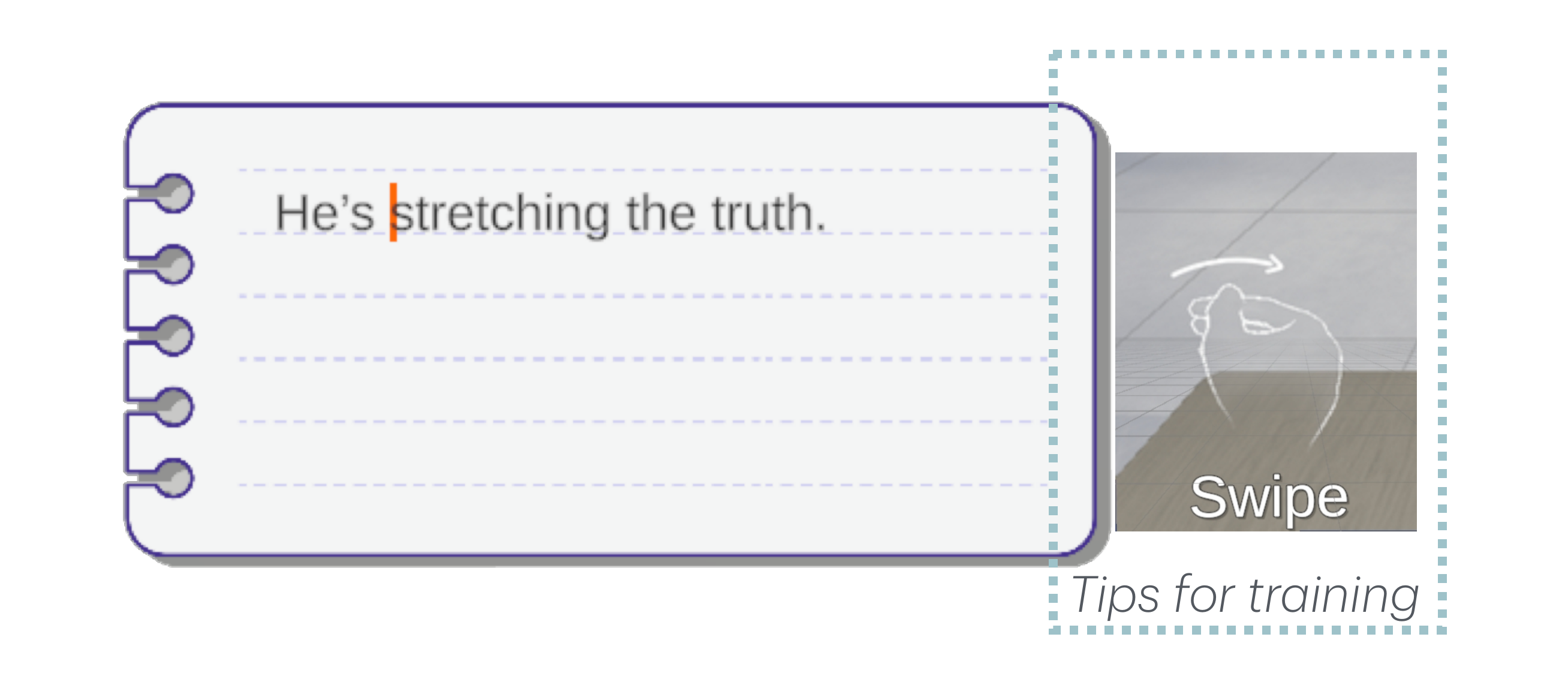}
    \caption{Screenshot from the training session in Study 1. The gesture instructions, shown in the blue dashed box, provide guidance on how to perform the required gestures (e.g., ``Swipe''). These tips were only available during the training session and were hidden during the formal sessions.}
    % \Description{Screenshot from User Study 3. The left screen displays a web browser titled 'Physical Features in Alps' with images and text about mountain ranges. The right screen shows a notes application with sample questions like 'Which mountain is the highest peak in the Alps, and what is its altitude?' Participants extract information from the web browser and paste it into the notes application.}
    \label{fig:study_1_screenshoot}
\end{figure}

\subsubsection{Study Design and Tasks}
\label{sec:Tasks}

In this study, participants followed instructions to use one of eight text editing tools to complete each command (e.g., ``Please Paste the Words'', ``Please Navigate the Caret''). The order of instructions was randomized within each group. In the Baseline condition, participants used ray casting to select a function and confirmed their choice with a pinch gesture. For caret navigation, they pinched to position the caret and confirmed it with the same gesture. Text selection involved pinching to start, dragging the caret to highlight text, and pinching again to confirm, replicating typical VR text editing techniques. The system automatically selected the relevant text, with participants only needing to execute the editing functions by pinching the corresponding button on the text editing toolbar. In the microGEXT condition, participants used specific gestures to complete commands (see Figure \ref{fig:system_feature}). For example, the \textit{Ring} gesture was used to execute the \textsc{Copy} function. Gesture tips were provided during the training session but not in the formal tasks (see Figure \ref{fig:study_1_screenshoot}), although researchers reminded participants of the gestures if needed. Similar to the Baseline, participants only needed to execute the gestures without selecting relevant text.

For caret navigation in microGEXT, participants positioned the caret with a pinch, followed by a \textit{Swipe} gesture to fine-tune placement, and confirmed by pressing their fingers together for two seconds. The same process applied to text selection. The long press gesture was detected automatically, ensuring precise timing to avoid errors. If users made a mistake, they had to reset the task with a pinch gesture. After each command, users received visual and auditory confirmation, and a loading bar signalled the preparation for the next command. Participants were instructed to rest their hands during this time. Each condition consisted of six rounds of eight instructions, with one training round and five formal rounds.

\subsection{Results}

%We used several measures to evaluate the usability and utility of both systems in this user study. For \textsc{Quantitative Performance}, we capture \textit{Edit Time across Commands}, referring to the time taken to complete each command; (2) \textit{Reattempts}, representing the number of times participants had to re-execute a command, either due to microGEXT misrecognizing the gesture or participants incorrectly performing a gesture or selecting the wrong menu item; (3) \textit{Average Edit Time}, as the time for each editing attempt, including both successful and reattempted executions; and (4) \textit{Learnability}, which was calculated based on the best-fitting learning curve, including the learning rate and learning gains.

To capture \textsc{Quantitative Performance}, we measured (1) \textit{Edit Time}, defined as the duration from when an editing command appeared to its successful execution. \rev{This measure included both initial and reattempted executions and was logged automatically by the system}; and (2) \textit{Reattempts}, defined as the number of times a command had to be re-executed due to either microGEXT misrecognition, incorrect gesture performance, or erroneous menu selection.

% \begin{table}[t]
% \caption{Comparison of Edit Times and Reattempt Counts between Baseline and microGEXT Conditions. Significant differences are highlighted with `*', `**', `***', and `****' indicating significance levels at $p < 0.05$, $p < 0.01$, $p < 0.001$, and $p < 0.0001$, respectively.}
% \centering
% \begin{tabular}{lcc}
% \toprule
% \textbf{Measure} & \textbf{Baseline (M ± SD)} & \textbf{microGEXT (M ± SD)} \\
% \midrule

% \multicolumn{3}{l}{\textbf{Task}} \\
% \textsc{Caret Navigation} & 9.52 ± 3.60 & 9.69 ± 2.02 \\

% \textsc{Range Selection} & 17.11 ± 6.82 & 13.75 ± 2.48 \\

% \textsc{Cut} & 5.73 ± 0.91 & \cellcolor{lightgray}4.95 ± 1.20 (*) \\

% \textsc{Paste} & 6.25 ± 2.04 & 5.80 ± 1.95 \\

% \textsc{Copy} & 5.38 ± 0.93 & 5.43 ± 1.49 \\

% \textsc{Undo} & 6.31 ± 1.58 & 5.68 ± 1.20 \\

% \textsc{Delete} & 7.24 ± 2.02 & \cellcolor{lightgray}4.99 ± 1.28 (***)\\

% \textsc{Select All} & 6.19 ± 1.46 & \cellcolor{lightgray}4.75 ± 1.55 (**) \\

% \midrule

% \textbf{Overall Edit Time} & 7.72 ± 1.72 & \cellcolor{lightgray}6.58 ± 0.78 (*)\\

% \textbf{Reattempts} & 0.32 ± 0.67 & \cellcolor{lightgray}2.25 ± 1.69 (****)\\

% \bottomrule
% \end{tabular}
% \end{table}

We also collected \textsc{Qualitative Feedback}, including: (1) \textit{User Experience}, measured using the short version of the User Experience Questionnaire (UEQ-S)~\cite{schrepp2017design}; (2) \textit{Perceived Workload}, evaluated using the NASA-TLX~\cite{hart_nasa-task_2006}; and (3) \textit{System Usability}, measured using the System Usability Scale (SUS)~\cite{Brooke1996SUSA}. We used the Shapiro–Wilks tests and Q-Q plots to check the normality distribution of the data. For normally distributed data, we used Welch’s t-test for two-level comparisons, while we used the Mann-Whitney U test for non-normally distributed data for two-level comparisons. To avoid distorting the statistical analysis, we removed outlier data points with an absolute Z-Score greater than 3.

\subsubsection{Quantitative Performance}

\paragraph{\textbf{Edit Time for Commands}}
We compared edit times between the Baseline and microGEXT conditions across various text editing tasks. For \textsc{Caret Navigation}, \textsc{Range Selection}, \textsc{Paste}, \textsc{Copy}, and \textsc{Undo}, no significant differences were observed ($p > 0.05$), indicating comparable performance between the two conditions.

In contrast, significant improvements were found in the \textsc{Cut}, \textsc{Delete}, and \textsc{Select All} tasks. For \textsc{Cut}, microGEXT was faster ($M = 4.95$, $SD = 1.20$) compared to the Baseline ($M = 5.73$, $SD = 0.91$), showing a significant difference ($U = 293.0$, $p = 0.0123$, $r = 0.398$). Similarly, for \textsc{Delete}, microGEXT ($M = 4.99$, $SD = 1.28$) outperformed the Baseline ($M = 7.24$, $SD = 2.02$), with a significant effect ($U = 332.0$, $p = 0.0004$, $r = 0.565$). Lastly, for \textsc{Select All}, microGEXT ($M = 4.75$, $SD = 1.55$) was faster than the Baseline ($M = 6.19$, $SD = 1.46$), showing significance ($U = 318.0$, $p = 0.0015$, $r = 0.505$).

\paragraph{\textbf{Overall Edit Time}}
Across all tasks, the microGEXT condition had a significantly shorter average edit time ($M = 6.58$, $SD = 0.78$) compared to the Baseline ($M = 7.72$,  $SD = 1.72$), as revealed by a Mann-Whitney U test ($U = 280.0$, $p = 0.0315$, $r = 0.342$).

\paragraph{\textbf{Reattempts}}
Participants required significantly more reattempts in the microGEXT condition ($M = 2.25$, $SD = 1.69$) than in the Baseline ($M = 0.32$, $SD = 0.67$), with a strong effect size ($U = 46.5$, $p < 0.0001$, $r = -0.657$).

\subsubsection{Qualitative Feedback}

\begin{table}[t]
\caption{Comparison of edit times, reattempt counts, UEQ, NASA-TLX, and SUS between Baseline and microGEXT. Significant differences are marked as `*', `**’, `***', and `****’ for $p < 0.05$, $p < 0.01$, $p < 0.001$, and $p < 0.0001$, respectively. Better results are indicated with (↑) and worse results with (↓).}
\centering
\begin{tabular}{lcc}
\toprule
\textbf{Measure} & \textbf{Baseline (M ± SD)} & \textbf{microGEXT (M ± SD)} \\
\midrule
\multicolumn{3}{l}{\textbf{Average Edit Times by Task}} \\
\textsc{Caret Navigation} & 9.52 ± 3.60 & 9.69 ± 2.02 \\
\textsc{Range Selection} & 17.11 ± 6.82 & 13.75 ± 2.48 \\
\textsc{Cut} & \textbf{5.73 ± 0.91} (↓) & \textbf{4.95 ± 1.20} (*) (↑) \\
\textsc{Paste} & 6.25 ± 2.04 & 5.80 ± 1.95 \\
\textsc{Copy} & 5.38 ± 0.93 & 5.43 ± 1.49 \\
\textsc{Undo} & 6.31 ± 1.58 & 5.68 ± 1.20 \\
\textsc{Delete} & \textbf{7.24 ± 2.02} (↓) & \textbf{4.99 ± 1.28} (***) (↑) \\
\textsc{Select All} & \textbf{6.19 ± 1.46} (↓) & \textbf{4.75 ± 1.55} (**) (↑) \\
\textbf{Overall Edit Time} & \textbf{7.72 ± 1.72} (↓) & \textbf{6.58 ± 0.78} (*) (↑) \\
\midrule
\textbf{Reattempts} & \textbf{0.32 ± 0.67} (↑) & \textbf{2.25 ± 1.69} (****) (↓) \\
\midrule
\multicolumn{3}{l}{\textbf{UEQ}} \\
Pragmatic Quality & 1.4 ± 1.12 & 1.35 ± 0.99 \\
Hedonic Quality & \textbf{-0.05 ± 1.47} (↓) & \textbf{1.96 ± 0.70} (***) (↑) \\
Overall Experience & \textbf{0.675 ± 1.12} (↓) & \textbf{1.66 ± 0.70} (**) (↑) \\
\midrule
\multicolumn{3}{l}{\textbf{NASA-TLX}} \\
Mental Demand & 3.55 ± 1.73 & 3.85 ± 1.98 \\
Physical Demand & 3.95 ± 1.47 & 3.1 ± 1.80 \\
Temporal Demand & 2.85 ± 1.42 & 3.15 ± 1.69 \\
Performance & 3.7 ± 1.45 & 3.0 ± 1.45 \\
Effort & 3.95 ± 1.57 & 4.0 ± 1.56 \\
Frustration & 2.65 ± 1.57 & 2.25 ± 1.21 \\
\midrule
\multicolumn{3}{l}{\textbf{SUS}} \\
Overall Usability & \textbf{69.38 ± 17.26} (↑) & \textbf{64.88 ± 18.58} (↓) \\
\bottomrule
\end{tabular}
\end{table}

\paragraph{\textbf{User Experience Questionnaire (UEQ)}}
The UEQ-Short results assessed Pragmatic Quality, Hedonic Quality, and Overall User Experience. For Pragmatic Quality, no significant difference was found between the Baseline ($M = 1.4$, $SD = 1.12$) and microGEXT ($M = 1.35$, $SD = 0.99$; $U = 212.50$, $p = 0.7439$, $r = 0.063$). However, for Hedonic Quality, microGEXT significantly outperformed the Baseline ($M = 1.96$, $SD = 0.70$ vs. $M = -0.05$, $SD = 1.47$; $U = 40.50$, $p < 0.0001$ , $r = 0.798$). Similarly, for Overall User Experience, microGEXT ($M = 1.66$, $SD = 0.70$) was significantly better than the Baseline ($M = 0.675$, $SD = 1.12$; $U = 100.00$, $p = 0.0070$, $r = 0.500$).

\paragraph{\textbf{NASA Task Load Index (NASA-TLX)}}
No significant differences were found between the Baseline and microGEXT conditions across all six workload subscales: Mental Demand, Physical Demand, Temporal Demand, Performance, Effort, and Frustration ($p > 0.05$ for all comparisons). For example, Mental Demand scores were similar ($M = 3.55$, $SD = 1.73$ for Baseline; $M = 3.85$, $SD = 1.98$ for microGEXT; $U = 184.50$, $p = 0.6800$), as were Physical Demand ($M = 3.95$, $SD = 1.47$ for Baseline; $M = 3.1$, $SD = 1.80$ for microGEXT; $U = 263.50$, $p = 0.0829$) and other subscales.

\paragraph{\textbf{System Usability Scale (SUS)}}
The SUS scores, which measure system usability, showed no significant difference between the Baseline ($M = 69.38$, $SD = 17.26$) and microGEXT ($M = 64.88$, $SD = 18.58$; $U = 227.00$, $p = 0.4728$, $r = 0.135$).

\paragraph{\textbf{Qualitative Comments}} Participants appreciated the system’s gesture recognition. For instance, \textit{``the gesture recognition is very fast and accurate''} [P1]. However, [P4] pointed out that learning the gestures took time and suggested the need for a dedicated gesture memory process to help users familiarize themselves. [P5] described the overall experience as smooth but mentioned, \textit{``micro-adjustments are convenient, but error tolerance is too low''}, especially during fine adjustments. [P7] and [P9] also found that the cursor was too sensitive, which sometimes led to mistakes during use. [P12] found the system innovative, stating that \textit{``(microGEXT is) fun and new, providing strong control''}, but [P14] commented that the gestures for actions like copy and paste were \textit{``unintuitive''}, requiring extra effort to memorize. [P13] noted that prolonged use of gestures, particularly swiping, caused hand discomfort. 

However, despite these concerns, many participants [P5, P18, P19] found that after becoming familiar with the system, microGEXT significantly reduced hand fatigue, saying \textit{``...after getting used to it, microGEXT is less tiring''} [P19], making it more efficient than traditional methods, like Baseline.

\section{User Study 2: Evaluating microGEXT for Precise and Rapid Text Range Selection}

\rev{Study 2 focuses on text range selection, a task that often requires both precision and speed in VR. Accordingly, we examine whether microGEXT provides relative advantages for short text range selections, where a minimal swipe may be more efficient than maintaining a pinch, but incurs higher time costs for longer ranges due to the required mode-switching step (\textbf{H3}). Meanwhile, because microgestures minimize physical effort and provide a compact interaction style, they may still be perceived as more comfortable and less fatiguing overall, which could result in higher usability and user experience ratings compared to the Baseline (\textbf{H4}).}

\subsection{Method}

\subsubsection{Baseline}

In the Baseline condition, users could select text character-by-character during the selection process. To begin, they controlled a ray to point at the starting position of the desired text. After positioning the ray, they pressed and held the trigger button on the controller, moving the ray to extend the selection over the intended text snippets. Once the caret reached the end of the selection, users released the trigger button to complete the selection. To confirm their selection, they pointed the ray at the ``Confirm'' button located on the left side of the interface and pressed the trigger again to finalize the task. As in many VR systems, a circular cursor appeared on the text panel to indicate the ray’s current position. If users made an error during the selection, however, they were required to restart the task from the beginning, repeating the entire process. This interaction method demanded high precision and was prone to errors, often resulting in frequent task restarts.

\subsubsection{Participants and Apparatus}

All previous participants ($N=20$) from Study 1 took part in User Study 2 in the same room, following the completion of the final questionnaire in Study 1 and a 5-minute break. \rev{The study protocol was also approved by the Cambridge University Engineering Department Ethics Committee (\#452).}

\subsubsection{Study Design and Tasks}

\begin{figure*}
    \centering
    \includegraphics[width=\linewidth]{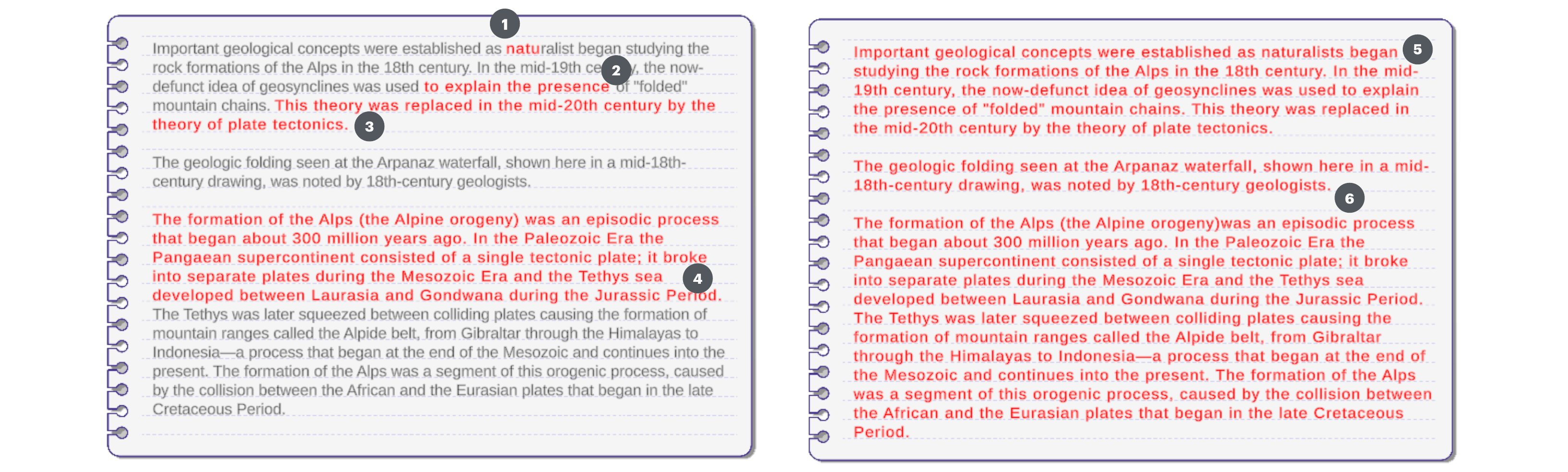}
    \caption{Example tasks used in our study 2: (1) Four characters, (2) Four words, (3) One sentence, (4) Two sentences, (5) One paragraph, (6) Two paragraphs.}
    \label{fig:study2_task}
\end{figure*}

We followed the same six tasks as described by Song et al.~\cite{song_exploring_2024}, which were initially adapted from Goguey et al.~\cite{Goguey_improving_2018}. Participants were required to select a target text snippet in VR as fast and accurately as possible using both conditions: Baseline and microGEXT. The target text snippets varied from six lengths:

\begin{itemize}
\item \textit{Four Characters (Four Char.)}: Selecting 4 characters from a 10-character word.
\item \textit{Four Words}: Selecting 4 consecutive words.
\item \textit{Sentence (Sent.)}: Selecting a complete sentence.
\item \textit{Two Sentences (Two Sent.)}: Selecting 2 consecutive sentences.
\item \textit{Paragraph (Para.)}: Selecting an entire paragraph.
\item \textit{Two Paragraphs (Two Para.)}: Selecting 2 consecutive paragraphs.
\end{itemize}

% \begin{figure*}
%     \centering
%     \begin{minipage}[t]{0.5\linewidth}
%         \centering
%         \includegraphics[width=0.7\linewidth]{Figures/Study2_command_edit_time.pdf}
%         \caption{Caption for the edit time}
%         \label{fig:nasa_tlx}
%     \end{minipage}%
%     \hfill
%     \begin{minipage}[t]{0.5\linewidth}
%         \centering
%         \includegraphics[width=0.3\linewidth]{Figures/Study2_edit_time.pdf}
%         \caption{Caption for the edit time}
%         \label{fig:sus}
%     \end{minipage}
% \end{figure*}

The key difference between the text selection tasks in User Study 1 and User Study 2 for the microGEXT condition was the introduction of a mode-switching mechanism for the left wrist, similar to the approaches used by Song et al.~\cite{song_exploring_2024} and Song et al.~\cite{song_efficient_2022}. In the microGEXT condition, users activated the mode switch menu by performing a thumb-up gesture. Once activated, the mode selection process was controlled via wrist rotation, with a vertical ray extending from the palm to guide the selection of the minimum selection unit. The mode aligned with the ray was highlighted in purple, providing visual feedback to the user. To confirm the selection, users relaxed their hands by spreading their four fingers. A visual change in the mode canvas, displayed in front of the user’s view, indicated the successful mode switch. The system then dynamically adjusted the minimum selection unit based on the selected mode, allowing for a smooth transition between different interaction states.

The text panel parameters were primarily based on those from Song et al.~\cite{song_exploring_2024}, with adjustments made to fit our experimental design. The panel was sized at 996px $\times$ 683px, with the text displayed within a designated area of 896px $\times$ 615px. We used the LiberationSans SDF font with white text, and the target text snippet was highlighted in red. The formatted text consisted of four paragraphs, totalling approximately 194 words. We selected a font size of 24 and a line spacing of 7.3, with left alignment. The text spanned 21 lines in total, including line breaks, with the longest line containing around 122 characters and the median line length at approximately 62 characters. To ensure readability, we conducted in-lab pilot tests with four participants, all of whom reported that the text was clear and easy to read, without any readability issues.

% For the text selection task, users can pinch first to put the caret in 

\subsubsection{Procedure}

It is similar to our first user study; participants were given a tutorial video and informed about the task of user study 2 in the following formal experiment. Then, participants also have one round of training sessions to familiarize themselves with the system. Following this, participants can have a short break before conducting the formal user study. Then, participants completed the formal trials for each condition, following the experimental design described in the previous section. The same questionnaires as Study 1 were given right after the completion of a condition, followed by a short break and a post-studies questionnaire assessing their experience (i.e., ease of use, presence, final rating) and perceived fatigue for both Study 1 and 2. Consequently, the overall study comprised a total of 1,200 trials, calculated as 2 (System) $\times$ 6 (Instruction) $\times$ 5 (Round) $\times$ 20 (Participant). The whole study would take approximately 40 minutes.

\subsection{Results}

In User Study 2, we used similar measures to evaluate the \textsc{Quantitative Performance} and \textsc{Qualitative Feedback}. Specifically, we collected: (1) Edit Time; (2) User Experience; (3) Perceived Workload; and (4) System Usability.

\subsubsection{Quantitative Performance}

% \begin{table}[t]
% \caption{Comparison of Average Edit Times and Overall Edit Time between Baseline and microGEXT Conditions Across Different Tasks. Significant differences are highlighted with `***' and `****' indicating significance levels at $p < 0.001$ and $p < 0.0001$, accordingly.}
% \centering
% \begin{tabular}{lcc}
% \toprule
% \textbf{Measure} & \textbf{Baseline (M ± SD)} & \textbf{microGEXT (M ± SD)} \\
% \midrule
% \multicolumn{3}{l}{\textbf{Task}} \\
% \textsc{Four Char.} & 28.04 ± 19.49 & 19.15 ± 4.43 \\

% \textsc{Four Words} & 21.23 ± 16.97 & 19.58 ± 4.75 \\

% \textsc{Sentence} & 21.24 ± 18.83 & 15.17 ± 3.21 \\

% \textsc{Two Sentences} & 16.98 ± 9.59 & 16.07 ± 3.90 \\

% \textsc{Paragraph} & \cellcolor{lightgray}9.96 ± 5.67 (***) & 14.82 ± 4.64 \\

% \textsc{Two Paragraphs} & \cellcolor{lightgray}10.38 ± 3.94 (****) & 15.51 ± 3.29 \\

% \midrule
% \textbf{Overall Edit Time} & 17.85 ± 11.42 & 16.69 ± 3.08 \\

% \bottomrule
% \end{tabular}
% \end{table}

\paragraph{\textbf{Average Edit Time}}
The average edit times for six tasks were compared between the Baseline and microGEXT conditions. For shorter text range selection tasks (\textsc{Four Char.}, \textsc{Four Words}, \textsc{Sentence}, and \textsc{Two Sentences}), no significant differences were observed ( $p > 0.05$ ), with both conditions demonstrating comparable performance. For example, in the \textsc{Four Char.} task, the Baseline mean was $M = 28.04$, $SD = 19.49$, while microGEXT had $M = 19.15$, $SD = 4.43$. Similarly, in the \textsc{Sentence} task, the Baseline mean was $M = 21.24$, $SD = 18.83$, compared to microGEXT ($M = 15.17$, $SD = 3.21$).

In contrast, significant differences were observed for longer text range selection tasks. For the \textsc{Paragraph} task, microGEXT was significantly slower ($M = 14.82$ , $SD = 4.64$ ) compared to the Baseline ($M = 9.96$, $SD = 5.67$; $U = 62.0$, $p = 0.0002$, $r = -0.590$). Similarly, in the \textsc{Two Paragraphs} task, microGEXT ($M = 15.51$, $SD = 3.29$) was significantly slower than the Baseline ($M = 10.38$, $SD = 3.94$; $U = 40.0$, $p < 0.0001$, $r = -0.684$).

\paragraph{\textbf{Overall Edit Time}}
When comparing the overall edit time across all tasks, the Baseline condition ($M = 17.85$, $SD = 11.42$) and microGEXT ($M = 16.69$, $SD = 3.08$) showed no significant difference ($U = 132.0$, $p = 0.0679$, $r = -0.291$).

\subsection{Qualitative Feedback}

\begin{table}[t]
\caption{Comparison of average and overall edit times, UEQ, NASA-TLX, and SUS scores between Baseline and microGEXT. Significant differences are marked as *, **, ***, and **** for $p < 0.05$, $p < 0.01$, $p < 0.001$, and $p < 0.0001$, respectively. Better results are indicated with (↑) and worse results with (↓).}
\centering
\begin{tabular}{lcc}
\toprule
\textbf{Measure} & \textbf{Baseline (M ± SD)} & \textbf{microGEXT (M ± SD)} \\
\midrule
\multicolumn{3}{l}{\textbf{Average Edit Times by Task}} \\
\textsc{Four Char.} & 28.04 ± 19.49 & 19.15 ± 4.43 \\
\textsc{Four Words} & 21.23 ± 16.97 & 19.58 ± 4.75 \\
\textsc{Sentence} & 21.24 ± 18.83 & 15.17 ± 3.21 \\
\textsc{Two Sentences} & 16.98 ± 9.59 & 16.07 ± 3.90 \\
\textsc{Paragraph} & \textbf{9.96 ± 5.67} (***) (↑) & \textbf{14.82 ± 4.64} (↓) \\
\textsc{Two Paragraphs} & \textbf{10.38 ± 3.94} (****) (↑) & \textbf{15.51 ± 3.29} (↓) \\
\textbf{Overall Edit Time} & 17.85 ± 11.42 & 16.69 ± 3.08 \\

\midrule
\multicolumn{3}{l}{\textbf{UEQ Scores}} \\
Pragmatic Quality & \textbf{0.26 ± 1.55} (↓) & \textbf{1.5 ± 1.04} (**) (↑) \\
Hedonic Quality & \textbf{-0.55 ± 1.49} (↓) & \textbf{1.96 ± 0.69} (***) (↑) \\
Overall Experience & \textbf{-0.14 ± 1.31} (↓) & \textbf{1.73 ± 0.69} (***) (↑) \\

\midrule
\multicolumn{3}{l}{\textbf{NASA-TLX Scores}} \\
Mental Demand & \textbf{4.5 ± 1.88} (↓) & \textbf{2.9 ± 1.17} (**) (↑) \\
Physical Demand & \textbf{5.0 ± 1.75} (↓) & \textbf{3.65 ± 1.46} (*) (↑) \\
Temporal Demand & 3.6 ± 1.67 & 3.15 ± 1.50 \\
Performance & 3.55 ± 1.64 & 2.65 ± 1.50 \\
Effort & \textbf{4.45 ± 1.76} (↓) & \textbf{3.2 ± 1.20} (*) (↑) \\
Frustration & \textbf{4.3 ± 1.81} (↓) & \textbf{2.2 ± 1.15} (*) (↑) \\

\midrule
\multicolumn{3}{l}{\textbf{SUS Scores}} \\
Overall Usability & \textbf{55.88 ± 24.69} (↓) & \textbf{70.42 ± 13.77} (*) (↑) \\

\bottomrule
\end{tabular}
\end{table}

\paragraph{\textbf{User Experience Questionnaire (UEQ)}}
The UEQ-Short results for Study 2 revealed significant differences in all dimensions. For Pragmatic Quality, microGEXT ($M = 1.5$, $SD = 1.04$) scored significantly higher than the Baseline ($M = 0.26$, $SD = 1.55$; $U = 104.00$ ,  $p = 0.0096$, $r = 0.480$). Hedonic Quality showed a highly significant improvement with microGEXT ($M = 1.96$, $SD = 0.69$) outperforming the Baseline ($M = -0.55$, $SD = 1.49$; $U = 33.50$,  $p < 0.0001$ , $r = 0.833$). For Overall User Experience, microGEXT ($M = 1.73$, $SD = 0.69$) also significantly surpassed the Baseline ($M = -0.14$, $SD = 1.31$; $U = 44.00$, $p < 0.0001$, $r = 0.780$).

\paragraph{\textbf{NASA Task Load Index (NASA-TLX)}}
The NASA-TLX results indicated significant reductions in workload with microGEXT across several subscales. Mental Demand was significantly lower for microGEXT ($M = 2.9$,  $SD = 1.17$) compared to the Baseline ($M = 4.5$, $SD = 1.88$; $U = 300.00$, $p = 0.0062$,  $r = 0.500$). Physical Demand also decreased significantly with microGEXT ($M = 3.65$,  $SD = 1.46$) versus the Baseline ($M = 5.0$, $SD = 1.75$; $U = 293.50$, $p = 0.0103$, $r = 0.468$). Similarly, Effort was significantly lower for microGEXT ($M = 3.2$, $SD = 1.20$) compared to the Baseline ($M = 4.45$, $SD = 1.76$; $U = 281.00$,  $p = 0.0261$, $r = 0.405$), and Frustration was markedly reduced with microGEXT ($M = 2.2$, $SD = 1.15$) versus the Baseline ($M = 4.3$, $SD = 1.81$; $U = 330.50$, $p = 0.0004$, $r = 0.653$).

No significant differences were observed for Temporal Demand ($M = 3.15$, $SD = 1.50$ for microGEXT vs. $M = 3.6$, $SD = 1.67$ for Baseline;  $U = 231.00$, $p = 0.3987$, $r = 0.155$) or Performance ($M = 2.65$, $SD = 1.50$ for microGEXT vs. $M = 3.55$, $SD = 1.64$ for Baseline; $U = 268.00$, $p = 0.0600$, $r = 0.340$).

\paragraph{\textbf{System Usability Scale (SUS)}}
The SUS results showed a significant improvement in system usability for microGEXT ($M = 70.42$, $SD = 13.77$) compared to the Baseline ($M = 55.88$,   $SD = 24.69$; $U = 104.50$, $p = 0.0101$, $r = 0.478$).

\paragraph{\textbf{{Qualitative Comments}}} Many appreciated the microGEXT’s precision, for example, [P2] noting that \textit{``(microGEXT) allows (me) for more precise control of text selection compared to the default system.''} However, [P10] noted that \textit{``the current selection status on the left-hand panel wasn’t clearly visible,''} suggesting it needed better visual clarity.

Gesture memorization and sensitivity were common concerns. [P3] remarked, \textit{``the system is easier to use than the default mode, but gestures require practice and are easily forgotten.''} Similarly, [P16] emphasized the need for sensitivity adjustments, as the current settings could lead to occasional misrecognition of gestures.

The left-hand panel was praised for its adaptability, as [P7] highlighted that \textit{``...it helps adapt to different text selection scenarios.''} But some suggested improvements in fluidity [P15] and smoother switching functions noted by [P11], while \textit{``microGEXT worked well for large text blocks, switching between functions could be smoother.''}

% \begin{table}[t]
% \caption{Comparison of Ease of Use, Preference, Presence, and Perceived Fatigue between Baseline and microGEXT conditions. Significant differences are highlighted.}
% \centering
% \begin{tabularx}{\linewidth}{X X X}
% \toprule
% \textbf{Measure} & \textbf{Condition} & \textbf{M (SD)} \\
% \midrule

% \textbf{Ease of Use} \\
% \rowcolor{lightgray} & Baseline & 4.05 (1.67) \\
% \rowcolor{lightgray} & microGEXT & 5.60 (0.99) \\
% \midrule

% \textbf{Preference} \\
% \rowcolor{lightgray} & Baseline & 4.15 (1.42) \\
% \rowcolor{lightgray} & microGEXT & 6.00 (0.79) \\
% \midrule

% \textbf{Presence} \\
% \rowcolor{lightgray} & Baseline & 4.80 (1.54) \\
% \rowcolor{lightgray} & microGEXT & 6.10 (1.02) \\
% \midrule

% \textbf{Perceived Fatigue} \\
% \rowcolor{lightgray} & Baseline & 12.70 (3.13) \\
% \rowcolor{lightgray} & microGEXT & 10.60 (2.11) \\
% \bottomrule
% \end{tabularx}
% \end{table}

\subsection{Post-studies Questionnaires}

We provided a post-studies questionnaire to evaluate participants' (1) Perceived Fatigue, evaluated by using Borg Rating of Perceived Exertion (RPE) Scale (Borg 6--20)~\cite{williams2017borg}, where 6 means ``no exertion at all (rest)'' and 20 represents maximal exertion, meaning the person is pushing themselves to their absolute physical limit; (2) Ease of Use; (3) Presence; and (4) Overall Preference, regarding the use of these two systems in both studies. The last three metrics were assessed via a single 7-point Likert scale, where 1 indicated ``not at all'' and 7 indicated ``very much.''

% \begin{figure}
%     \centering
%     \includegraphics[width=\linewidth]{Figures/crop_microGEXT_afterall.pdf}
%     \caption{The figure illustrates the comparison between the Baseline and microGEXT conditions across four metrics: Presence, Preference, Ease of Use, and Perceived Fatigue. The box plots show the median (the horizontal line)), the mean (`x'), the first and third quartile (the box) and the minimum and maximum (the whiskers). The dot signs (`◦') represent outliers. Significant differences between conditions are annotated above the bars, with `*', `**' and `***' indicating significance levels at $p < 0.05$, $p < 0.01$ and $p < 0.001$, respectively.} 
%     % \Description{Box plots comparing Baseline and microGEXT across four metrics: Presence, Preference, Ease of Use, and Perceived Fatigue. microGEXT shows higher ratings for Presence, Preference, and Ease of Use, while Perceived Fatigue is lower. Box plots show the median, interquartile range, and outliers for both conditions.}
%     \label{fig:afterall}
% \end{figure}

\begin{table}[t]
\caption{Comparison of Ease of Use, Preference, Presence, and Perceived Fatigue between Baseline and microGEXT Conditions. Significant differences are marked with `*', `**', and `***', indicating significance levels at $p < 0.05$, $p < 0.01$, and $p < 0.001$, respectively. Better results are indicated with (↑) and worse results with (↓).}
\centering
\begin{tabular}{lcc}
\toprule
\textbf{Measure} & \textbf{Baseline (M ± SD)} & \textbf{microGEXT (M ± SD)} \\
\midrule

\textbf{Ease of Use} & \textbf{4.05 ± 1.67} (↓) & \textbf{5.60 ± 0.99} (**) (↑)\\
\midrule
\textbf{Preference} & \textbf{4.15 ± 1.42} (↓) & \textbf{6.00 ± 0.79} (***) (↑)\\
\midrule
\textbf{Presence} & \textbf{4.80 ± 1.54} (↓) & \textbf{6.10 ± 1.02} (**) (↑)\\
\midrule
\textbf{Perceived Fatigue} & \textbf{12.70 ± 3.13} (↓) & \textbf{10.60 ± 2.11} (*) (↑)\\

\bottomrule
\end{tabular}
\end{table}

\paragraph{\textbf{Perceived Fatigue}}
Participants reported significantly less perceived fatigue with microGEXT ($M = 10.60$, $SD = 2.11$) than with the Baseline ($M = 12.70$, $SD = 3.13$; $U = 273.50$, $p = 0.0450$, $r = 0.684$).

\paragraph{\textbf{Ease of Use}}
The microGEXT condition ($M = 5.60$, $SD = 0.99$) was rated significantly easier to use than the Baseline ($M = 4.05$,  $SD = 1.67$;  $U = 86.50$, $p = 0.0015$,  $r = 0.216$).

\paragraph{\textbf{Presence}}
MicroGEXT provided a stronger sense of presence ($M = 6.10$,  $SD = 1.02$) compared to the Baseline ($M = 4.80$, $SD = 1.54$; $U = 100.50$, $p = 0.0055$,  $r = 0.251$).

\paragraph{\textbf{Preference}}
Participants strongly preferred the microGEXT condition ($M = 6.00$, $SD = 0.79$) over the Baseline ($M = 4.15$, $SD = 1.42$; $U = 52.00$, $p = 0.00004$, $r = 0.130$), with a highly significant difference.

\section{User Study 3: Exploring microGEXT for Open-Ended Information Gathering in Web Browsing and Note-Taking}

We conducted User Study 3 to allow participants to fully experience microGEXT in an open-ended information-gathering scenario in web browsing and note-taking. 

\subsection{Method}

\subsubsection{Participants and Apparatus}

We invited all previous participants to participate in User Study 3, conducted at least one day after completing the first two studies. Half of the participants ($N=10$, 5 male, 5 female) from the earlier user studies were willing to participate. The age range of the participants was between 17 and 26 years ($M = 23.5, SD = 2.68$), and the VR familiarity range was between 1 and 6 ($M = 4.4, SD = 1.58$). The study took place in the same room using the same apparatus as the previous sessions.

\subsubsection{Procedure}

Participants in User Study 3 engaged in an open-ended scenario focused on information gathering through web browsing and note-taking, without specific tasks or wrong actions. The study was conducted in a virtual environment, where participants viewed two screens simultaneously: the left screen displayed a web browser with information about various travel-related topics (e.g., mountain locations, opening hours, altitudes, etc.), and the right screen showed a notes application with sample questions (e.g., what is the altitude of the mountain). Participants were required to find the relevant answers from the web browser and then add the proper information into the notes application. \rev{The study protocol was also approved by the Cambridge University Engineering Department Ethics Committee (\#452).}

At the beginning of the study, participants were asked to revisit the video tutorial that explained how to use both the microGEXT and Baseline conditions for text editing in browsers and note applications. Following the video, participants are allowed to complete a training session to familiarize themselves with the microGEXT system. The training session involved browsing a website and transferring information to the notes application, mainly by copying and pasting editions. After the training, participants proceeded to the formal round, which consisted of two conditions: one using microGEXT and the other using the Baseline. Each condition involved interacting with a different website to gather and transfer information. After completing the tasks in both conditions, participants filled out the same questionnaires as in the previous studies to assess their experience. The entire study took approximately 30 minutes to complete, and participants received an additional £5 as compensation for their time and effort.

\begin{figure}
    \centering
    \includegraphics[width=\linewidth]{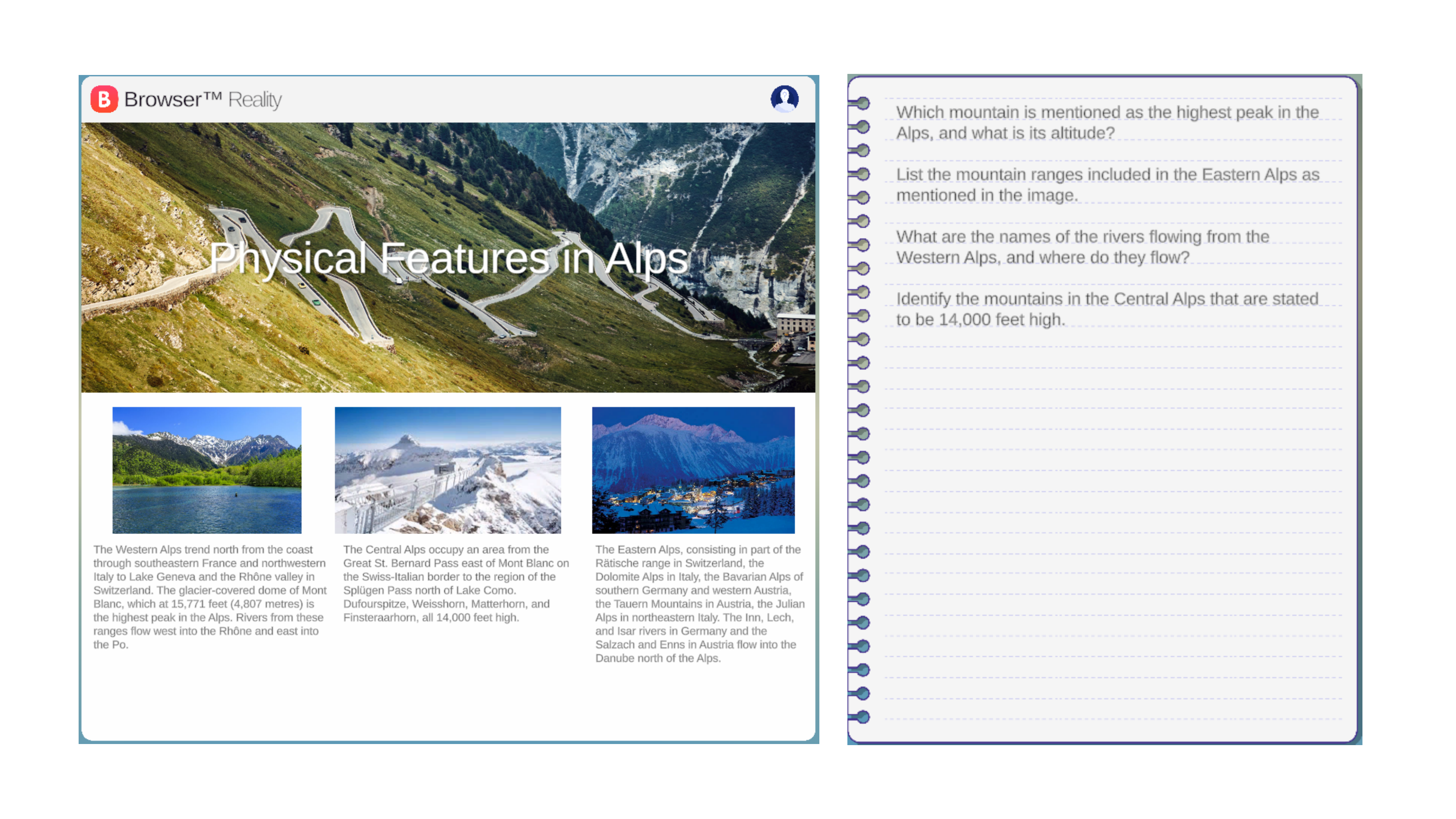}
    \caption{Screenshot from User Study 3, showing the scenario where participants performed information gathering and note-taking tasks. The left side of the screen displays a web browser with travel-related information about the Alps, and the right side shows a notes application with sample questions that participants needed to answer by extracting relevant data from the web page.}
    % \Description{Screenshot from User Study 3 showing a web browser titled 'Physical Features in Alps' on the left, and a notes application with sample questions on the right. Participants gather information from the browser and input it into the notes app.}
    \label{fig:study3_screenshot}
\end{figure}

\subsection{Results}

In Study 3, given the smaller group of participants, we focused primarily on subjective questionnaires and semi-structured interviews to gather \textsc{Qualitative Feedback} from participants. Specifically, we collected the feedback across three key areas: (1) User Experience; (2) Perceived Workload; and (3) System Usability. Considering that the scenarios were open-ended and reading speeds varied among individuals, we did not collect edit times during this study.

\subsubsection{Qualitative Feedback}

\paragraph{\textbf{User Experience Questionnaire (UEQ)}}
The Pragmatic Quality scores showed no significant difference between the Baseline ($ M = 1.25$, $SD = 1.31$) and microGEXT ($M = 1.9$, $SD = 0.59$; $U = 39.50$, $p = 0.4471$, $r = 0.210$). However, significant differences were observed for Hedonic Quality, where microGEXT ($M = 1.9$, $SD = 0.60$) outperformed the Baseline ($M = -0.425$, $SD = 1.31$; $U = 9.00$, $p = 0.0021$, $r = 0.820$). Similarly, Overall User Experience scores were significantly higher for microGEXT ($M = 1.9$, $SD = 0.51$) compared to the Baseline ($M = 0.4125$, $SD = 1.09$; $U = 14.50$, $p = 0.0080$, $r = 0.710$).

\paragraph{\textbf{NASA Task Load Index (NASA-TLX)}}
The NASA-TLX results revealed no significant differences between the Baseline and microGEXT conditions across all subscales ($p > 0.05$). For instance, Mental Demand was similar ($M = 3.6$, $SD = 1.84$ for Baseline; $M = 3.5$, $SD = 1.58$ for microGEXT; $U = 49.50$, $p = 1.0000$), as were Physical Demand ($M = 3.8$, $SD = 1.23$ for Baseline; $M = 3.6$, $SD = 1.65$ for microGEXT; $U = 56.00$, $p = 0.6700$) and Temporal Demand ($M = 3.1$, $SD = 1.10$ for Baseline; $M = 3.3$,  $SD = 1.25$  for microGEXT; $U = 44.00$, $p = 0.6682$). Performance ($U = 25.00$, $p = 0.0586$), Effort ($U = 50.50$, $p = 1.0000$), and Frustration ($U = 54.00$, $p = 0.7836$) also showed no significant differences.

\paragraph{\textbf{System Usability Scale (SUS)}}
The SUS results indicated no significant difference in system usability scores between the Baseline ($M = 69$, $SD = 21.51$) and microGEXT ($M = 74.5$, $SD = 13.17$; $U = 40.50$, $p = 0.4955$, $r = 0.190$).

\begin{table}[t]
\caption{Comparison of UEQ, NASA-TLX, and SUS scores between Baseline and microGEXT. Significant differences are marked with `**’ for $p < 0.01$. Better results are indicated with (↑) and worse results with (↓).}
\centering
\begin{tabular}{lcc}
\toprule
\textbf{Measure} & \textbf{Baseline (M ± SD)} & \textbf{microGEXT (M ± SD)} \\
\midrule

\textbf{UEQ} & & \\
Pragmatic Quality & 1.25 ± 1.31 & 1.9 ± 0.59 \\
Hedonic Quality & \textbf{-0.43 ± 1.31} (↓) & \textbf{1.9 ± 0.60} (**) (↑) \\
Overall Experience & {0.41 ± 1.09} (↓) & \textbf{1.9 ± 0.51} (**) (↑) \\

\midrule
\textbf{NASA-TLX} & & \\
Mental Demand & 3.6 ± 1.84 & 3.5 ± 1.58 \\
Physical Demand & 3.8 ± 1.23 & 3.6 ± 1.65 \\
Temporal Demand & 3.1 ± 1.10 & 3.3 ± 1.25 \\
Performance & 2.1 ± 1.10 & 3.7 ± 1.95 \\
Effort & 3.6 ± 1.65 & 3.6 ± 1.51 \\
Frustration & 2.8 ± 2.04 & 2.0 ± 0.67 \\

\midrule
\textbf{SUS} & & \\
Overall Usability & 69.0 ± 21.51 & 74.5 ± 13.17 \\

\bottomrule
\end{tabular}
\end{table}

\paragraph{\textbf{{Qualitative Comments}}}

Participants in Study 3 highlighted both the strengths and areas for improvement of microGEXT, particularly in relation to its efficiency for continuous tasks. [P9] noted that \textit{``...for consecutive operations like copy-paste, microGEXT allows for quick and seamless completion of both tasks, whereas the default method requires separate actions through the panel, which reduces efficiency.''} Once familiar with the system, [P9] found microGEXT to be \textit{``simple and efficient for note-taking, provided the gesture recognition was accurate.''} [P19] echoed the sentiment of efficiency, remarking that \textit{``once familiar with the system, users can quickly perform the intended functions.''} However, they pointed out that the accuracy of the gesture recognition for locking gestures could be improved, and mentioned that the gesture memorization process \textit{``requires some time, making microGEXT slightly more challenging compared to the default method.''} [P18] appreciated the speed and convenience of microGEXT, saying that it was \textit{``much faster than traditional methods and quite convenient.''} However, they observed that \textit{``(microGEXT) sometimes selects unnecessary spaces when recognizing paragraphs or sentences.''} 

\section{Discussion, Limitations, and Future Work}

A key finding across all studies is the significant reduction in physical demand and fatigue when using microGEXT. Traditional VR input methods, especially those that rely on large arm movements, are prone to causing users fatigue over time, as exemplified by the \emph{gorilla arm} syndrome~\cite{palmeira2024quantifying,hincapi2014consumed}. microGEXT addresses this challenge effectively through small, precise hand movements that reduce the physical strain typically associated with text editing tasks in VR. The NASA-TLX results from Studies 1 and 2 support this, showing that microGEXT consistently lowered physical and mental demand, while also reducing frustration, particularly in tasks requiring high precision like text selection and deletion. \rev{In future work, fatigue evaluation could benefit from computational modeling approaches such as NICER~\cite{li_nicer_2024} or adaptive interventions like AlphaPIG~\cite{li_alphapig_2025}, which provide predictive and real-time measures of physical strain in XR.}

\rev{Another dimension concerns gesture memorability. Prior elicitation studies~\cite{chan_user_2016} and large-scale surveys of gesture sets~\cite{villarreal-narvaez_brave_2024} have shown that as the number of gestures increases, users experience difficulties in recall and consistency. Recent work on transferable microgestures across contexts~\cite{chaffangeon_caillet_microgesture_2025} reinforces the need to balance ergonomic feasibility with cognitive load. Our studies suggest that microGEXT’s compact set is memorable, but expanding beyond eight commands may require hybrid designs or mnemonic aids to prevent usability degradation.}

While the recognition system achieved high accuracy, \rev{our dataset was limited in size and diversity, raising potential concerns about generalization to broader user populations. Adding new gestures would also necessitate retraining, limiting scalability. Future work could leverage data augmentation or transfer learning to expand coverage without exhaustive new data collection. Cross-population studies involving users with different cultural or ergonomic backgrounds would further strengthen generalizability.}

\rev{The current system relies solely on microgestures, which highlights both strengths and limitations. While effective for short-range editing, performance slowed in longer-range selection tasks (Study 2). This suggests the need for hybrid modalities that combine microgestures with other inputs such as voice commands or contextual menus, allowing users to flexibly adapt to task complexity. Exploring such multimodal synergies could preserve efficiency while maintaining low fatigue.}

Finally, \rev{we see important opportunities in deployment studies. Lab-based tasks capture controlled performance metrics but miss appropriation, adaptation, and long-term use patterns that emerge in everyday workflows. Inspired by recent work on extended deployment in XR~\cite{biener2022quantifying}, future research could examine microGEXT in authentic productivity contexts to uncover barriers, adoption strategies, and sustained benefits.}

In summary, microGEXT highlights the potential of microgestures for efficient and low-fatigue VR text editing, while also surfacing broader design and methodological challenges around fatigue evaluation, memorability, dataset scalability, and multimodal integration.

\section{Conclusion}

This paper introduced \emph{microGEXT}, a microgesture-based system for text editing in VR, designed to mitigate well-known challenges such as user fatigue, precision limitations, and social accessibility. We evaluated its usability, efficiency, and overall user experience through three controlled studies.

Study~1 ($N=20$) compared microGEXT with a commercially inspired Baseline ray-casting + pinch menu system, showing significant reductions in edit time for commands such as \textsc{Cut}, \textsc{Delete}, and \textsc{Select All}, alongside improved user experience. Study~2 ($N=20$) examined text range selection, where microGEXT performed comparably to the Baseline for short selections while reducing physical and mental demand in longer tasks. Study~3 ($N=10$) demonstrated that microGEXT was intuitive and fatigue-free in open-ended workflows, supporting seamless task switching during extended use. Our findings demonstrate that microgestures can provide a lightweight yet powerful alternative to conventional VR text editing techniques. By reducing fatigue and enhancing user satisfaction while retaining most of the performance of mid-air interfaces, microGEXT highlights the potential of subtle input for everyday VR productivity. Future work should explore refining recognition accuracy, extending the gesture vocabulary, and investigating long-term adoption in real-world scenarios, including hybrid modalities that combine microgestures with complementary input channels.

\section*{Acknowledgments}
Xiang Li is supported by the China Scholarship Council (CSC) International Cambridge Scholarship (No. 202208320092). Wei He is supported by the HKUST(GZ) Research Studentship. Per Ola Kristensson is supported by the EPSRC (grant EP/W02456X/1).

% {\appendices
% \section*{Proof of the First Zonklar Equation}
% Appendix one text goes here.
% You can choose not to have a title for an appendix if you want by leaving the argument blank
% \section*{Proof of the Second Zonklar Equation}
% Appendix two text goes here.}

\bibliographystyle{IEEEtran}
\bibliography{microgesture}

% \vfill\eject

% \newpage

% \vspace{11pt}

\vspace{-33pt}
\begin{IEEEbiography}[{\includegraphics[width=1in,height=1.25in,clip,keepaspectratio]{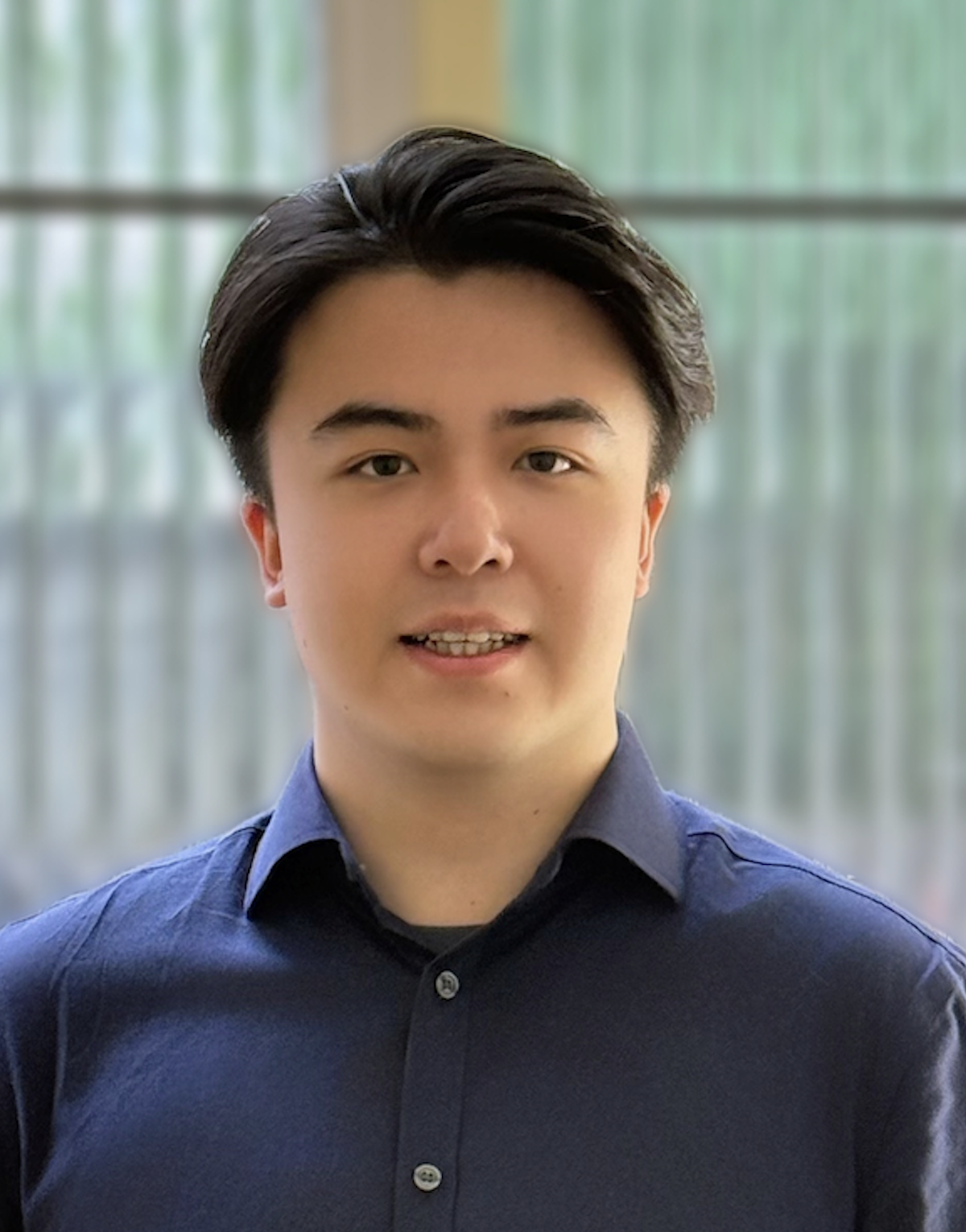}}]{Xiang Li}
is a PhD student in the Intelligent Interactive Systems group at the University of Cambridge and a Student Fellow at The Leverhulme Centre for the Future of Intelligence. He obtained his dual BSc degrees from the University of Liverpool and Xi'an Jiaotong-Liverpool University in 2022. Previously, he worked at Monash University, Carnegie Mellon University, the Institute Polytechnique de Paris (Télécom Paris), and the Hong Kong University of Science and Technology (Guangzhou). 
\end{IEEEbiography}

\vspace{11pt}

\vspace{-33pt}
\begin{IEEEbiography}[{\includegraphics[width=1in,height=1.25in,clip,keepaspectratio]{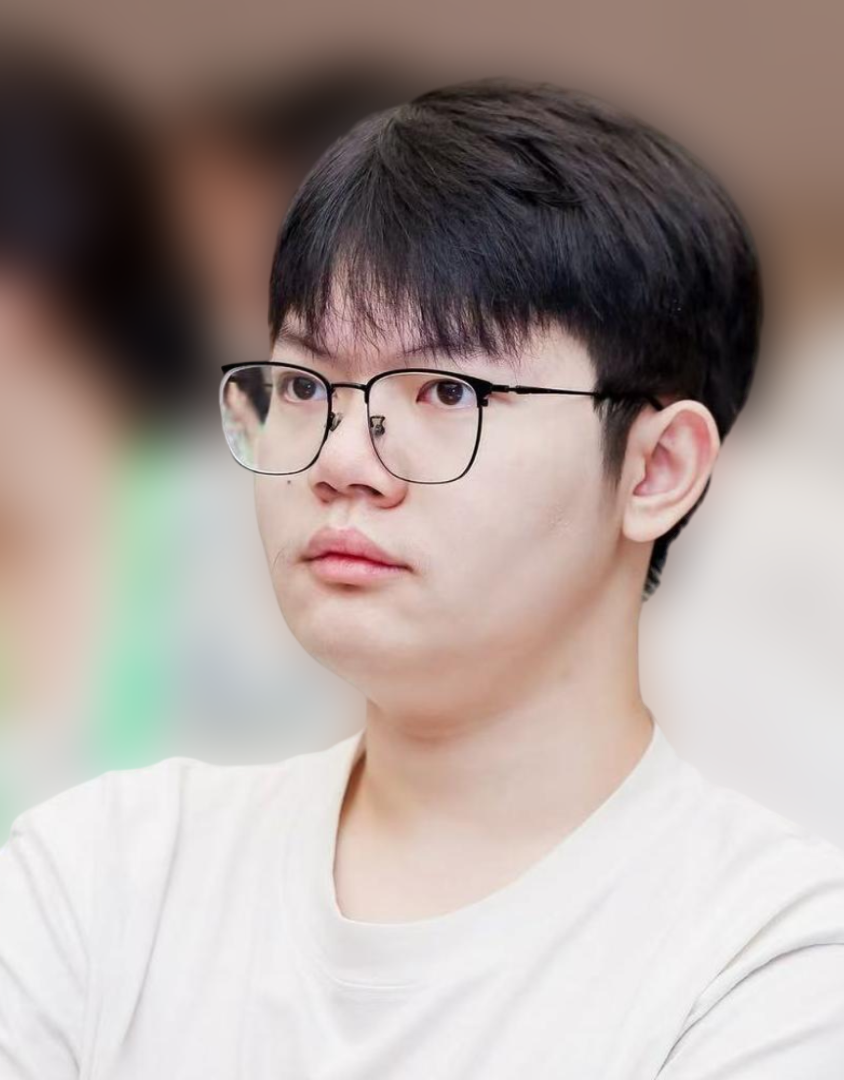}}]{Wei He}
is a PhD student in Urban Governance and Design at the Hong Kong University of Science and Technology (Guangzhou). Before this, he worked as a research assistant in the Department of Industrial and Systems Engineering at The Hong Kong Polytechnic University and at the Society Hub of the Hong Kong University of Science and Technology (Guangzhou) in 2023. He earned his BSc in Vehicle Engineering from Hunan University in 2022.
\end{IEEEbiography}

\vspace{11pt}

\vspace{-33pt}
\begin{IEEEbiography}[{\includegraphics[width=1in,height=1.25in,clip,keepaspectratio]{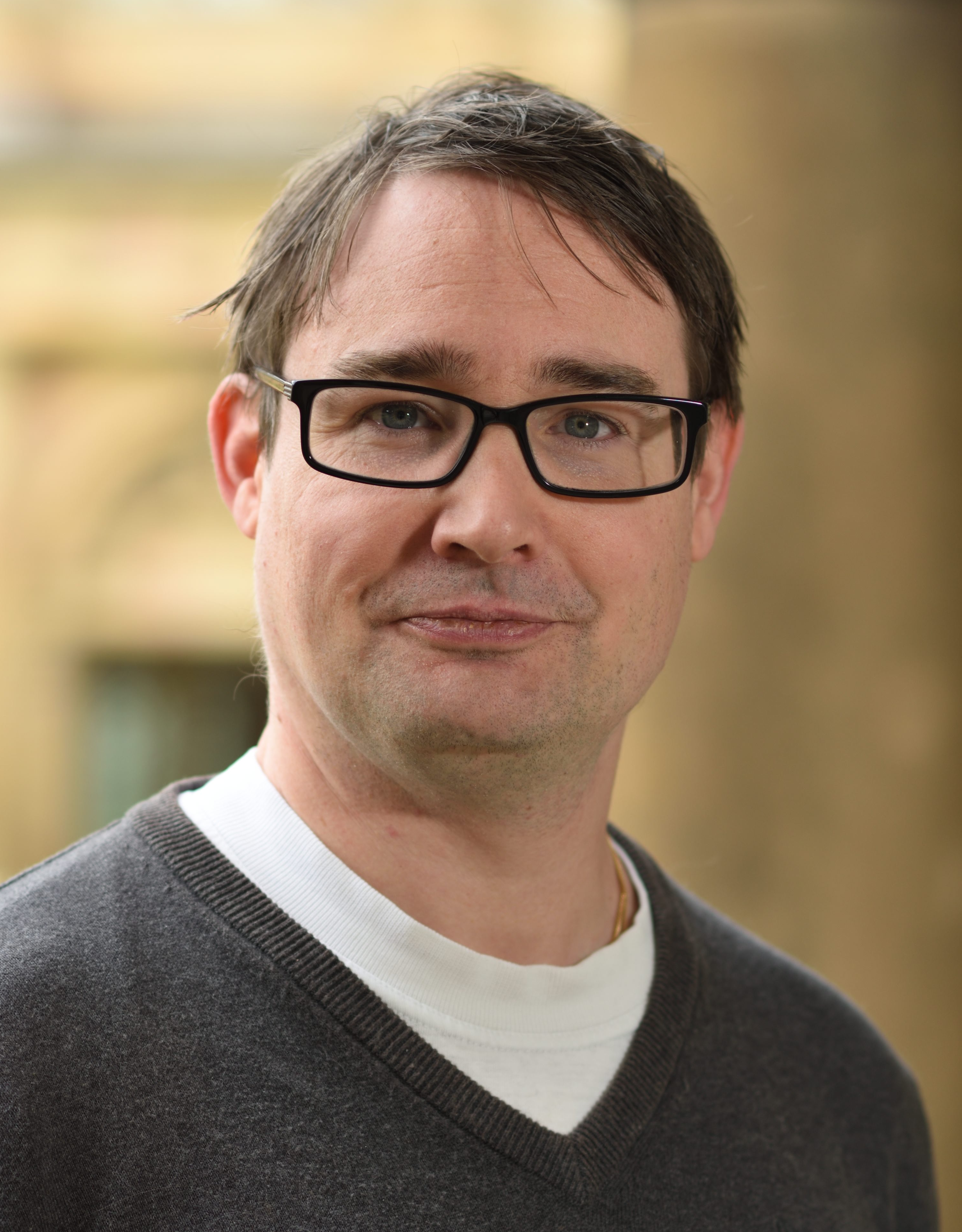}}]{Per Ola Kristensson}
is a Professor of Interactive Systems Engineering in the Department of Engineering at the University of Cambridge and a Fellow of Trinity College. He is also a co-founder and co-director of the Centre for Human-Inspired Artificial Intelligence at the University of Cambridge. He is an Associate Editor of ACM Transactions on Computer-Human Interaction and ACM Transactions on Interactive Intelligent Systems and serves as a Steering Committee Member for ACM CHI.
\end{IEEEbiography}

% \vfill

\end{document}